# Problems of search and pursuit of unmanned aerial vehicles using the game-theoretic approach

Oleg Malafeyev [1, a]    Kun Zhang [2, b]

[1] *Department of Modeling of Socio-economic Systems, Saint-Petersburg State University, 7/9 Universitetskaya nab., 199034 , Russian Federation .*

[2] *Department of Mathematical Theory of Microprocessor Control Systems, Saint-Petersburg State University,7/9 Universitetskaya nab., 199034, Russian Federation.*

[a] malafeyevoa@mail.ru

[b] st113011@student.spbu.ru

Corresponding author：st113011@student.spbu.ru

**Abstract**. Unmanned aerial vehicles (UAVs) have become increasingly prevalent in various domains, ranging from military operations to civilian applications. However, the proliferation of UAVs has also given rise to concerns regarding their potential misuse and security threats. As a result, the search and pursuit of UAVs have become crucial tasks for law enforcement agencies and security organizations.

In this paper, we use a game theoretic approach to explore the problem of searching for and pursuing submarines and translate the problem into a UAV search and pursuit problem. Game theory provides a mathematical framework for modeling and analyzing strategic interactions among multiple decision makers. By applying game theoretic principles to the search and pursuit problem, we aim to improve the effectiveness of UAV detection and capture strategies.

We begin by formulating the problem as a game, where the UAV represents the evader, and the search and pursuit team represents the pursuers. Each player's objective is to optimize their own utility while considering the actions and strategies of the other players.

By leveraging game theory, we can gain insights into the optimal decision-making strategies for both the UAV and the pursuers, leading to improved search and pursuit outcomes and enhanced security in the face of UAV threats.

# CONTENTS

# DYNAMIC MODELS OF INSPECTIONS

Currently, the method of mathematical modeling has gained wide popularity, with the benefit of constructing and studying mathematical models for the purpose of analyzing and forecasting various processes in natural, technical, economic, and other sciences [1-7]. The application of mathematical theory can be used to prevent illegal actions. Terrorists and members of drug cartels use modern means of communication and transportation in their activities. There is a need to conduct inspection measures to prevent the spread of their actions. To organize successful countermeasures, it is also necessary to use modern technical means and an apparatus for optimizing the use of resources, and to design dynamic models of inspections. Let us consider the following situation. An interceptor ship equipped with sonar detected the periscope of a submarine, which immediately disappeared in an unknown direction. It is necessary to intercept the submarine in the shortest possible time. Let us assume that the interceptor ship does not know the exact speed of the submarine. However, a discrete set of speeds is known, one of which is the actual speed of the submarine. Next, we will refer to the interceptor ship as P and the submarine as E, respectively.

First, let us present an algorithm for finding the search time under conditions where the speed of the escaping submarine is unknown to the interceptor. Suppose that the speed of the interceptor is much greater than the speed of the escaping submarine. At the initial moment of time of detection, the ship accurately determines the location of the submarine. Thus, the distance between him and the escaping submarine, denoted by $D_0$, is known. To find the interception time, it is necessary to determine the trajectory along which the interceptor ship should move. We introduce the polar coordinate system ρ and φ in such a way that the pole, point O, is located at the point of detection of the submarine, and the polar axis passes through the point where the interceptor ship is located. Then, the dynamics of the escaping submarine are described by equations:

$$\dot{\rho}^E = v$$

$$\dot{\varphi}^E = 0$$

The pursuer does not know the speed v with certainty, but it is known that it is chosen from a discrete set $V^E$. The maximum possible speed of the pursuer ship is denoted by $V^P$. The pursuer can guarantee the capture by trying all elements of the set $V^E$. Initially, the ship assumes that the runaway has a speed $v_1 \in V^E$. To capture the submarine at time $t_0$, the pursuer begins moving at a speed of $V^P$ towards point O and continues until time $t_1$, at which point the players are at the same distance from point O, meaning that the equation is satisfied.

$$\rho_1^P = \rho_1^E$$

And

$$\int_{t_0}^{t_1} v_1 dt + V^P(t_1 - t_0) = D_0$$

From time $t_1$, the pursuer must move, selecting a speed such that they constantly remain at the same distance from point O as the fleeing ship. To achieve this, the speed of the intercepting ship is divided into two components: radial $V_\rho$ and tangential $V_\varphi$. The radial component is the speed at which the ship moves away from the pole, i.e.

$$V_\rho = \dot{\rho}$$

The tangential component is the linear rotational velocity with respect to the pole, i.e.

$$V_\varphi = \rho\dot\varphi$$

To make the encounter happen, the radial component of the pursuer's velocity is assumed to be equal to the velocity of the fugitive. Then, to find the trajectory of the pursuer, the system of differential equations must be solved:

$$\dot\rho = v_1$$

$$\dot\varphi^2 \rho^2 = (V^P)^2 - (v_1)^2$$

The initial conditions for this system will be:

$$\varphi(t^*) = 0$$

$$\rho(t_1) = v_1 t_1$$

Solving it, we find:

$$\varphi(t) = \frac{\sqrt{(V^P)^2 - (v_1)^2}}{v_1} \ln\frac{v_1 t}{v_1 t_1}$$

$$\rho(t) = v_1 t$$

Let's express time as a function of the polar angle:

$$t(\varphi) = t_1 \exp\left(\frac{v_1 \varphi}{\sqrt{(V^P)^2 - (v_1)^2}}\right)$$

Thus, the trajectory consists of linear segments and logarithmic spiral segments. In [2], it is proven that during movement along the spiral, the encounter will occur in a time not exceeding the time of passing one turn. Therefore, if the ship, having bypassed the turn of the spiral, does not find the submarine, then the initial assumption about the speed of the fleeing vessel was incorrect. Then the next speed $v_2 \in V^E$ is chosen. Thus, the fleeing vessel during time $t_2$ covered the distance $\rho_E(t_2) = v_2 t_2$, and the pursuer $\rho_P(t_2) = v_1 t_2$. If $\rho_P(t_2) > \rho_E(t_2)$, then the distance between the players will be equal to $D_2 = \rho_P(t_2) - \rho_E(t_2)$, and to find the moment of time $t_3$, it is necessary to solve the equation

$$\int_{t_2}^{t_3} v_2 dt + V^P(t_3 - t_2) = D_2$$

If $\rho_P(t_2) < \rho_E(t_2)$, then the distance between the players will be equal to $⟦D_2 = \rho_E(t_2) - \rho_P(t_2)$ and to find the time $t_3$ it is necessary to solve the equation

$$V^P(t_3 - t_2) - \int_{t_2}^{t_3} v_2 dt = D_2$$

After moving along a straight section, the pursuer moves along a spiral. To reduce the time, it is expedient for the pursuer to order the speed search in descending order. However, if this becomes known to the evader, he can move at a minimum speed, which will maximize the search time. Thus, the following game is obtained. The set of strategies for the submarine is the set of combinations of possible speeds $v$ of its movement and directions of movement $\alpha$. The set of strategies for the intercepting ship is the set of all possible permutations of elements $V^E$. The matrix of the resulting game consists of elements T, which represent the capture time.

Now suppose that the intercepting ship needs to detect n submarines, each of which requires $\tau_{ij}$ hours to capture. To carry out the interception, there are m boats, each of which is directed to a submarine. The matrix $A = \{\tau_{ij}\}$ is known, which represents the efficiency matrix of search operation for the i-th boat and j-th submarine. The task is to construct such an assignment plan $X = \{x_{ij}\}, i = 1..m, j = 1..n$, which minimizes the search time, while assigning each boat to search for no more than one submarine, and each submarine can be searched by no more than one boat. The values of $x_{ij}$ can only take two values:

$$x_{ij} = \begin{cases} 1, assigned\ i\ boat\ for\ j\ submarine \\ 0, assigned\ i\ boat\ for\ j\ submarine \end{cases}$$

Mathematical formulation of the optimal assignment problem

$$min\ z = min \sum_{i=1}^{m}\sum_{j=1}^{n} \tau_{ij} * x_{ij}$$

$$\sum_{i=1}^{m} x_{ij} \leq 1, j = 1..n$$

$$\sum_{j=1}^{n} x_{ij} \leq 1, i = 1..m$$

$$x_{ij} \geq 0$$

In order for the problem of optimal assignments to have an optimal solution, it is necessary and sufficient that the number of boats is equal to the number of submarines, i.e., $n = m$. Under this condition, the inequality constraints become equalities.

$$min\ z = min \sum_{i=1}^{n}\sum_{j=1}^{n} \tau_{ij} * x_{ij}$$

$$\sum_{i=1}^{n} x_{ij} = 1, j = 1..n$$

$$\sum_{j=1}^{n} x_{ij} = 1, i = 1..n$$

$$x_{ij} \geq 0$$

If $n \neq m$, then the assignment problem is unbalanced. Any assignment problem can be balanced by introducing the necessary number of dummy boats or submarines. The dual problem of the optimal assignments.

$$max\ \omega = max(\sum_{i=1}^{n} i + \sum_{i=1}^{n} i)$$

$$i + i \geq \tau_{ij}, i = 1..n, j = 1..n$$

Hungarian method for solving assignment problems.

In the original performance matrix, A, determine the minimum element in each row and subtract it from all other elements in the row.

In the matrix obtained in the first step, find the minimum element in each column, and subtract it from all other elements in the column. If a feasible solution is not obtained after steps 1 and 2, the following should be performed:

a. In the last matrix, draw the minimum number of horizontal and vertical lines across rows and columns to cross out all zero elements.

b. Find the minimum non-crossed out element and subtract it from all other non-crossed out elements and add it to all elements at the intersection of the lines drawn in the previous step.

c. If the new distribution of zero elements does not allow for a feasible solution, repeat step 2a. Otherwise, proceed to step 3.

The optimal assignments will correspond to the zero elements obtained in step 2.

Let's consider another case. Suppose that an interceptor ship sends n boats after a single submarine. The escaping submarine has a discrete set of speeds and directions of movement, and it needs to choose how to act to maximize the time of capture. In other words, the escaping submarine must choose the best course of action or the best behavioral strategy. Let's use decision theory. Each boat tries to intercept the submarine one at a time in a random order. Therefore, we have n steps. Let's say we are at step t. It is necessary to determine the probability of winning in case strategy t is chosen, assuming that it is better than all the previous ones, i.e. the probability that it is the best one at all. Let's denote this probability by $g_t$ In addition, let's define the probability that the last strategy will be the best, if we skip the first t strategies and then the escaping submarine uses the optimal strategy. Let's denote this probability by $h_t$. According to the principle of dynamic programming, the escaping submarine knows how to act optimally starting from step t+1. The optimal behavioral strategy is: if the strategy at step t is not better than all the previous ones, then it should be rejected; if it is indeed better among the first t, then we need to compare $g_t$ and $h_t$. If $g_t \geq h_t\_t$, then the escaping submarine chooses

If the n-1 strategy is worse than the n-2 strategy, then the n-2 strategy is the best among the first n-1 strategies. The probability

$$g_{n-2} = \frac{1}{n-1} * 0 + \frac{n-2}{n-1} * g_{n-1} = \frac{n-2}{n}$$

I

$h_t$ is a monotonically non-increasing function. Insert graphs of functions. According to the chosen behavior strategy, if the $g_{n-1} = \frac{n-1}{n}$. It follows from the second case that E skips the $n$--strategy. Then the chances of winning $h_{n-1} = \frac{1}{n}$. Means

$$h_{n-2} = \frac{1}{n-1} * \frac{n-1}{n} + \frac{n-2}{n-1} * \frac{1}{n} = \frac{(n-2)+(n-1)}{n*(n-1)}$$

$$h_t = \frac{t}{n}\left(\frac{1}{t} + \frac{1}{t+1} + \cdots + \frac{1}{n-1}\right)$$

$\frac{h_t}{g_t}$ for

$$\frac{h_t}{g_t} = \frac{1}{t} + \frac{1}{t+1} + \cdots + \frac{1}{n-1}$$

The number corresponding to the intersection point on the graph was found in [3] and is equal to $t = \frac{n}{e}$. In this case $h_t = g_t = \frac{t}{n} = \frac{1}{e}$, i.e. the probability of success for $n \to \infty$ is $\frac{1}{e} = 0{,}368$.

Using the Maple software package, several examples were solved.

**Example 1**. Let the initial distance between the pursuer and the fugitive be 200 kilometers. The fugitive chooses a speed from the set $V^E = \{8, 56, 78\}$ and a direction from the set $\alpha = \{23, 137, 182\}$. The maximum speed of the pursuer is $V^P = 100$км/ч. Then the set of fugitive strategies is:

$$(\alpha_1, v_1), (\alpha_1, v_2), (\alpha_1, v_3), (\alpha_2, v_1), (\alpha_2, v_2), (\alpha_2, v_3), (\alpha_3, v_1), (\alpha_3, v_2),$$

set of pursuer strategies:

$$(v_1, v_2, v_3), (v_1, v_3, v_2), (v_2, v_1, v_3), (v_2, v_3, v_1), (v_3, v_1, v_2), (v_3, v_2, v_1)$$

The resulting game matrix looks like this:

$$\begin{matrix}
11{,}7 & 11{,}7 & 819{,}8 & 4{,}71*10^6 & 29547 & 2{,}07*10^6 \\
3{,}62*10^7 & 2{,}08*10^{11} & 7{,}24*10^6 & 7{,}24*10^6 & 9{,}12*10^{10} & 1{,}82*10^{10} \\
2{,}53*10^{15} & 3{,}62*10^{13} & 2{,}53*10^{15} & 5{,}06*10^{14} & 3{,}17*10^{12} & 3{,}17*10^{12} \\
\\
1{,}1*10^5 & 1{,}1*10^5 & 7{,}71*10^6 & 4{,}43*10^{10} & 2{,}78*10^8 & 1{,}94*10^{10} \\
1{,}06*10^{41} & 6{,}07*10^{44} & 2{,}11*10^{40} & 2{,}11*10^{40} & 2{,}66*10^{44} & 5{,}32*10^{43} \\
1{,}3*10^{77} & 1{,}86*10^{75} & 1{,}3*10^{77} & 2{,}6*10^{76} & 1{,}63*10^{74} & 1{,}63*10^{74} \\
\\
4{,}09*10^6 & 4{,}09*10^6 & 2{,}86*10^8 & 1{,}64*10^{12} & 10^{10} & 7{,}19*10^{11} \\
1{,}71*10^{54} & 9{,}84*10^{57} & 3{,}42*10^{53} & 3{,}42*10^{53} & 4{,}31*10^{57} & 8{,}62*10^{56} \\
2{,}98*10^{101} & 4{,}26*10^{99} & 2{,}97*10^{101} & 5{,}95*10^{100} & 3{,}73*10^{98} & 3{,}73*10^{98}
\end{matrix}$$

The game can be solved using any method for solving matrix games.

We transform the topic into the search and pursuit between quadrotor UAVs. Modify the topic slightly

Let the distance between the fugitive UAV and the ground be 100 meters, the fugitive UAV selects a speed from the set $V^E=\{8,56,78\}$ as the X-axis speed, and selects from the set $\alpha=\{23,37,82\}$ A value as the Y-axis direction. The maximum speed of the chaser is $V^P=120$ m/min

Then the fugitive policy set is:

$$(\alpha_1,v_1),(\alpha_1,v_2),(\alpha_1,v_3),(\alpha_2,v_1),(\alpha_2,v_2),(\alpha_2,v_3),(\alpha_3,v_1),(\alpha_3,v_2)$$

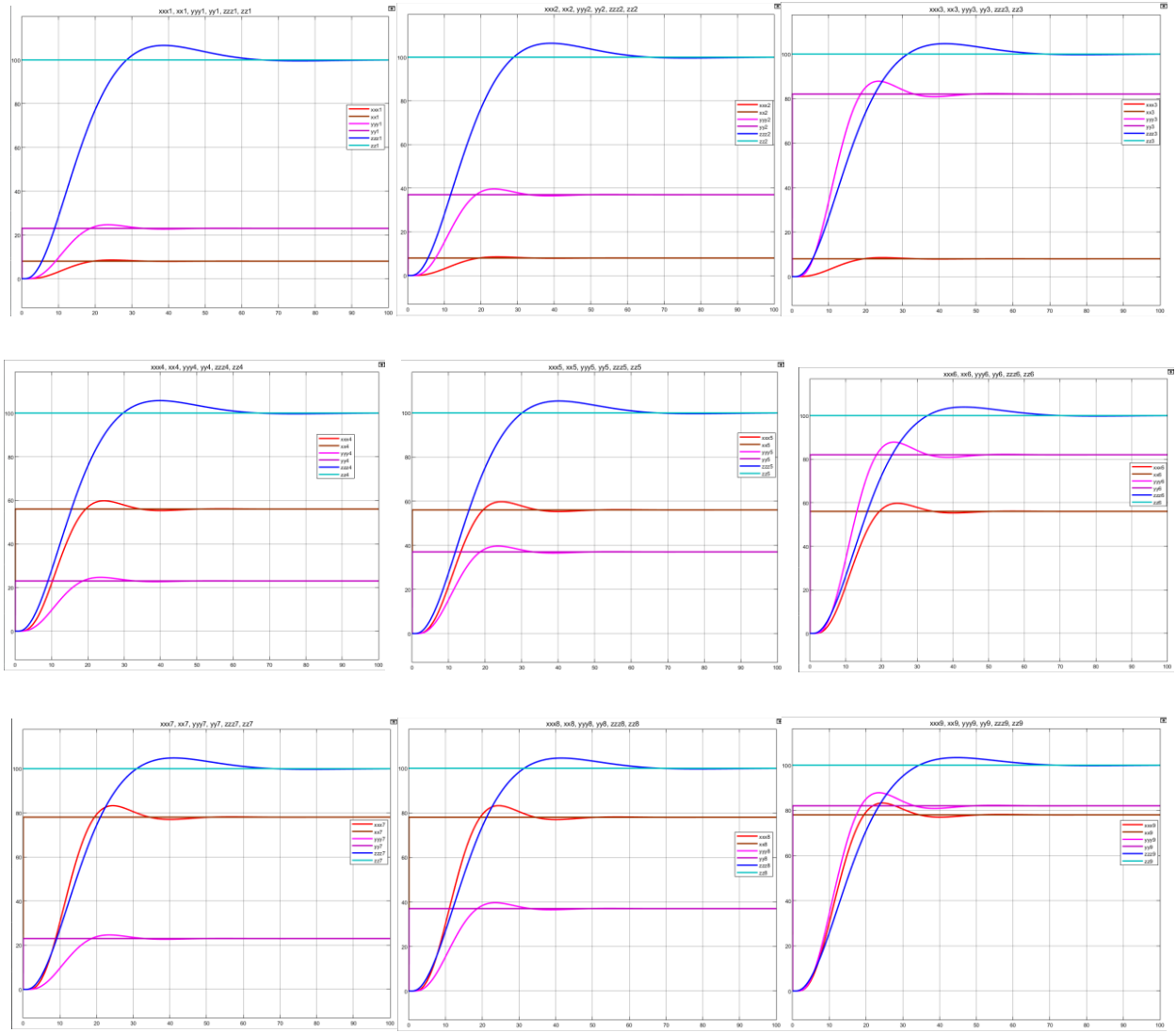

Figure.1.Nine strategies for simulating the runaway drone.

**Example 2**. Suppose a interceptor ship has detected 4 submarines. The initial distance to each of them is 100 kilometers, 200 kilometers, 50 kilometers, and 163 kilometers, respectively. The pursuer has 4 boats to catch the submarines. The maximum speed of each boat is 74 km/h, 90 km/h, 178 km/h, and 124 km/h, respectively. The first submarine moves along the straight line $\alpha_1=23$, at a speed of $v_1=23$ km/h, the second one - $\alpha_2=137$, $v_2=50$ km/h, the third one - $\alpha_3=187$, $v_3=67$ km/h, and the fourth one - $\alpha_4=50$, $v_4=70$ km/h. Then the matrix for the assignment problem looks as follows:

$$\begin{matrix} 1903 & 386 & 9{,}96 & 52 \\ 1{,}15*10^{71} & 6{,}4*10^{51} & 1{,}3*10^{34} & 1{,}89*10^{26} \end{matrix}$$

$$\begin{matrix} 5{,}6*10^{172} & 1{,}13*10^{90} & 2*10^{32} & 3{,}7*10^{51} \\ 2{,}4*10^{63} & 7{,}56*10^{26} & 1{,}28*10^{9} & 5{,}96*10^{14} \end{matrix}$$

The game can be solved using the Hungarian method.

We transform the topic into the search and pursuit between quadrotor UAVs. Modify the topic slightly

Suppose an intercepting quadcopter detects 4 intruding quadcopters. Chaser has 4 ships to chase the submarine. The maximum speed of each ship in XYZ axis is 74 km/h, 90 km/h, 178 km/h and 124 km/h respectively.

The first invasion quadrotor UAV, the maximum speed of the X-axis $v\_1=23$m/min, the maximum speed of the Y-axis $\alpha\_1=23$m/min, the height is 100 meters

The second invading quadrotor UAV has a maximum speed of X-axis $v\_2=50$m/min, a maximum speed of Y-axis $\alpha\_2=137$m/min, and a height of 200 meters.

The third invading quadrotor UAV, the maximum speed of the X-axis $v\_3=67$m/min, the maximum speed of the Y-axis $\alpha\_3=7$m/min, and a height of 50 meters

The fourth intrusion quadrotor UAV, the maximum speed of the X-axis $v\_4=70$m/min. Y-axis maximum speed $\alpha\_4=50$m/min, height 163 meters

matching matrix:

$$\begin{matrix} 0 & 0 & 1 & 0 \\ 1 & 0 & 0 & 0 \\ 0 & 1 & 0 & 0 \\ 0 & 0 & 0 & 1 \end{matrix}$$

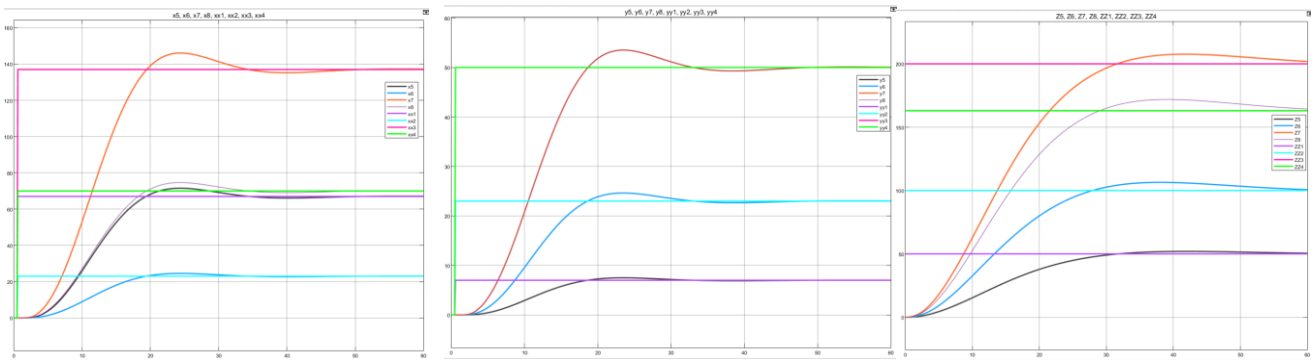

Figure.2.Simulate the X, Y, and Z axis value changes of the chaser and escaper drones.

**Example 3.** Suppose the initial distance between the pursuer and the evader was 50 kilometers. The evader chooses a velocity from the set $V^E = \{4,10,16\}$, and a direction from the set $\alpha = \{8,10,16\}$. The maximum speed of the pursuer is $V^P=80$ km/h. Then, the set of evader's strategies is:

$$(\alpha_1, v_1), (\alpha_1, v_2), (\alpha_1, v_3), (\alpha_2, v_1), (\alpha_2, v_2), (\alpha_2, v_3), (\alpha_3, v_1), (\alpha_3, v_2),$$

set of pursuer strategies:

$$(v_1, v_2, v_3), (v_1, v_3, v_2), (v_2, v_1, v_3), (v_2, v_3, v_1), (v_3, v_1, v_2), (v_3, v_2, v_1)$$

$$\begin{matrix} 0{,}73 & 0{,}73 & 1{,}6 & 6{,}75 & 2{,}6 & 5{,}79 \\ 1{,}5 & 6{,}2 & 0{,}92 & 0{,}92 & 5{,}28 & 3{,}32 \\ 4{,}83 & 2{,}19 & 4{,}83 & 3{,}03 & 1{,}18 & 1{,}18 \end{matrix}$$

$$\begin{matrix} 0{,}98 & 0{,}98 & 2{,}17 & 9{,}12 & 3{,}54 & 7{,}81 \\ 3{,}12 & 13{,}1 & 1{,}96 & 1{,}96 & 11{,}2 & 7{,}06 \\ 16{,}45 & 7{,}45 & 16{,}45 & 10{,}3 & 4{,}01 & 4{,}01 \end{matrix}$$

$$\begin{matrix} 1{,}33 & 1{,}33 & 2{,}93 & 12{,}3 & 4{,}78 & 10{,}56 \\ 6{,}64 & 27{,}95 & 4{,}17 & 4{,}17 & 23{,}96 & 15{,}04 \\ 55{,}99 & 25{,}37 & 55{,}99 & 35{,}14 & 13{,}65 & 13{,}65 \end{matrix}$$

The game can be solved by any method of solving matrix games.

We transform the topic into the search and pursuit between quadrotor UAVs. Modify the topic slightly

Let the distance between the fugitive UAV and the ground be 100 meters, the fugitive UAV selects a speed from the set V^E={4,10,16} as the X-axis speed, and selects from the set α={8,10,16} A value as the Y-axis direction. The maximum speed of the chaser is V^P=80 m/min

. Then the fugitive policy set is:

$$(\alpha_1, v_1), (\alpha_1, v_2), (\alpha_1, v_3), (\alpha_2, v_1), (\alpha_2, v_2), (\alpha_2, v_3), (\alpha_3, v_1), (\alpha_3, v_2)$$

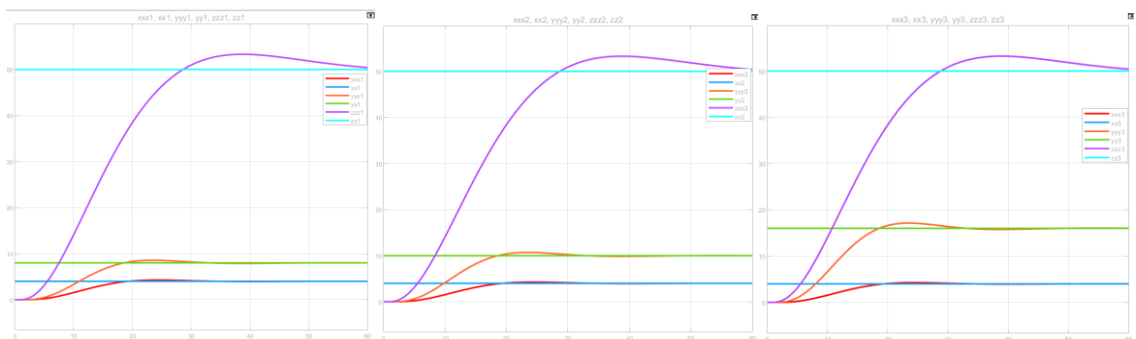

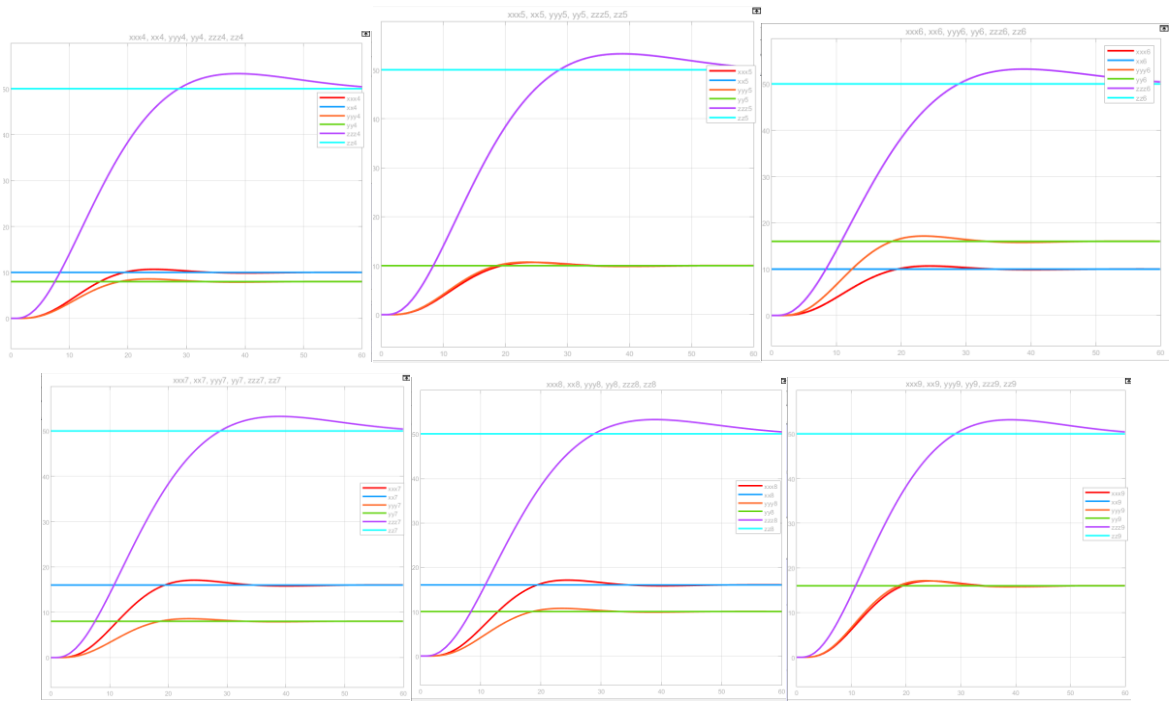

Figure.3.Nine strategies for simulating the runaway drone.

**Example 4**. Suppose a ship-interceptor has detected 4 submarines. The initial distance to each of them is 30 kilometers, 11 kilometers, 62 kilometers, and 8 kilometers, respectively. The pursuer has 4 boats to catch the submarines. The maximum speed of each boat is 60 km/h, 65 km/h, 95 km/h, and 105 km/h, respectively. The first submarine moves along the line $\alpha_1=7$ with a speed of $v1=7$ km/h, the second - $\alpha2=11$, $v2=11$ km/h, the third - $\alpha3=30$, $v3=30$ km/h, and the fourth - $\alpha4=44$, $v4=44$ km/h. Then the matrix for the assignment problem looks as follows:

$$\begin{pmatrix} 1{,}02 & 0{,}89 & 0{,}49 & 0{,}43 \\ 1{,}2 & 0{,}96 & 0{,}37 & 0{,}3 \\ 2{,}3*10^7 & 3{,}9*10^6 & 10758 & 3519{,}7 \\ 3{,}14*10^{19} & 2{,}75*10^{16} & 5{,}6*10^8 & 3{,}54*10^7 \end{pmatrix}$$

The game can be solved using the Hungarian method.

We transform the topic into the search and pursuit between quadrotor UAVs. Modify the topic slightly

Suppose an intercepting quadcopter detects 4 intruding quadcopters. Chaser has 4 ships to chase the submarine. The maximum speed of each ship in XYZ axis is 31 m/min, 12 m/min, 63 m/min and 9 m/min respectively.

The first invasion quadrotor UAV, the maximum speed of the X-axis v_1=7m/min, the maximum speed of the Y-axis α_1=7m/min, the height is 30 meters

The second invading quadrotor UAV has a maximum speed of X-axis v_2=11m/min, a maximum speed of Y-axis α_2=11m/min, and a height of 11 meters.

The third invading quadrotor UAV, the maximum speed of the X-axis v_3=30m/min, the maximum speed of the Y-axis α_3=30m/min, and a height of 62 meters

The fourth intrusion quadrotor UAV, the maximum speed of the X-axis v_4=44m/min. Y-axis maximum speed α_4=44m/min, height 44 meters

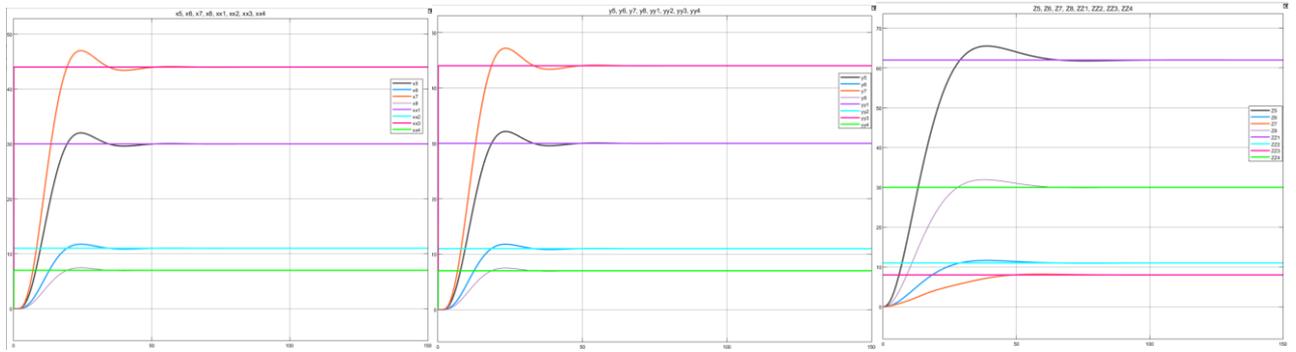

Figure.4. Simulate the X, Y, and Z axis value changes of the chaser and escaper drones.

**Example 5**. Suppose a ship-interceptor has detected 5 submarines. The initial distance between him and the escaping submarines is the same and is 20 km. After detection, each submarine goes in an unknown direction and at a different speed. The interceptor must intercept all escaping submarines. To do this, he sequentially intercepts each submarine. The first submarine moves along $\alpha_1$=18, with a speed of $v_1$=18 km/h, the second - $\alpha_2$=33, $v_2$=33 km/h, the third - $\alpha_3$=38, $v_3$=38 km/h, the fourth - $\alpha_4$=45, $v_4$=45 km/h, the fifth - $\alpha_4$=50, $v_4$=50 km/h. The speed of the ship-interceptor is 250 km/h. Using the Mong python chicken weight we c

suppose

Drone Interceptor detects 5 drones.

drone interceptor

The five intruders are quadcopter drones with heights of 20 m 40 m 60 m 80 m 100 m.

The speed of the first submarine along the straight-line X-axis v_1=120 m/min, y-axis speed α_1=20m/min,

The second X-axis velocity v_2=90 m/min, y-axis velocity α_2=50m/min,

The third X-axis speed v_3=70 m/min Y-axis speed α_3=70m/min,

The fourth X-axis velocity v_4=50 m/min, -y-axis velocity α_4=90m/min,

The fifth X-axis velocity v_5=20 m/min- y-axis velocity α_5=120m/min, .

The ship interceptor has a speed of 250 m/min.

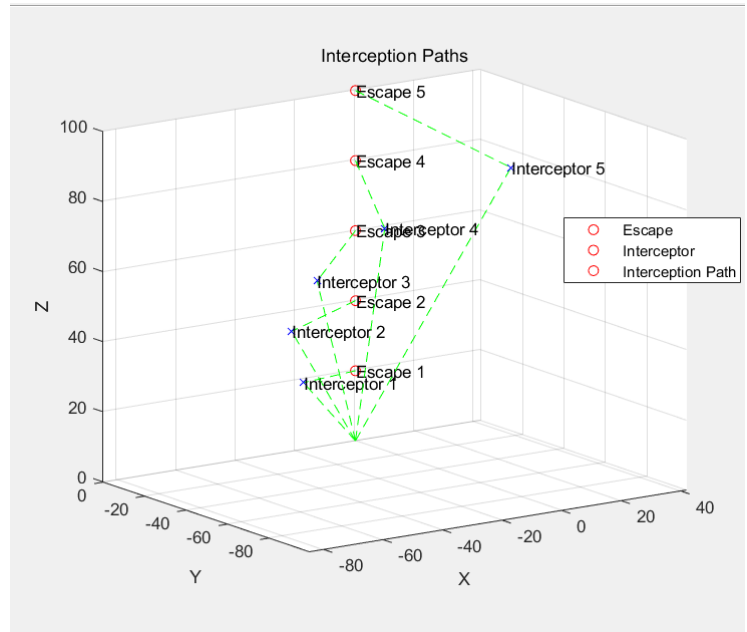

Figure.5. Quadrotor drone tracking path

# SEARCH FOR A MOVING OBJECT UNDER DIFFERENT INFORMATION CONDITIONS

## INTRODUCTION

Research on the problem of search began intensively during World War II. Later, search theory methods found application in solving various other problems [8], [9]. Two monographs on the theory of search for moving objects can be distinguished, devoted to various aspects of search theory [10], [11]. All developments in the theory of object search can be conditionally divided into three groups: discrete search, continuous search, and search game problems. The first group includes works devoted to the search for an object with a finite or countable set of states. The second group includes works on the search for an object whose set of positions forms a certain (finite or infinite) area in the plane or in the search space. The tasks of search in the first and second groups are discussed, in which the object being sought does not counteract the search system. Studies that consider search problems in conditions of counteraction form the third group, namely search game problems. The following articles can be distinguished from the works of the first group. In [12], Staroverova O.V. presents a search algorithm that minimizes the average search time under the condition that the probability $\alpha_i$ does not depend on the cell number and the search ends only when the pursuer detects the fleeing object. Kelin investigated the possibility of solving the problem of optimal distribution of search efforts using a Markov model, when the fleeing object E moves randomly. He considered two different cases:

1 The movement of the evader does not depend on their location, and the pursuer does not receive any information about the location of the evader during the search process.

2 The movement of the evader depends on their last location, and the search system knows the location of the evader in the previous moment at each moment in time.

In both cases, it is assumed that the search system knows the probabilistic law of the evader's movement. [13]

Koopman considered the problem of search on a surface, where the sought objects are uniformly distributed and their course angles are unknown (with equal probability in $[0, 2\pi]$). Assuming that the searcher P moves at a constant speed, he determined the average number of targets that enter a given circle centered on P per unit time at an arbitrary angle $\gamma \epsilon [0, 2\pi)$ and at a specified angle

$\gamma \in [\gamma', \gamma' + d\gamma']$, where $\gamma$ is the angle between the searcher's velocity vector and the line connecting P and E [14]. In [15], Kupman solved the problem of optimal distribution of given search efforts, maximizing the probability of detecting a stationary target with known a priori distribution of its location and an exponential conditional detection function. This was the first general result obtained in this field.

Lapshin presented Kupman's theory and provided examples showing how to use it to solve specific problems[16]. Charnes and Cooper [17] formalized the problem of optimal distribution of search efforts as a convex programming problem. Mac-Kuen and Miller[18] investigated the search problem, in which it is necessary to decide whether to start the search or not, and after starting the search, whether to continue or stop the search. They obtained a general functional equation. Pozner[19] solved the two-stage (preliminary and final) search problem for a lost satellite.

Dubrovin and Siro tin[20] solved the problem of determining the average time of finding an escaping object in a rectangular search area, if the initial locations of the pursuer and the escaping object in the specified area follow a uniform probability distribution and their courses are arbitrary. The sought object tries to leave the search area after being detected by the pursuer

One of the first solved game search problems is as follows: an airplane is searching for a submarine that is passing through a strait with variable width and a sufficiently large length; the submarine cannot remain submerged for the entire crossing time; it is necessary to determine the optimal distribution of search efforts for the airplane and the probability density distribution of the submarine's submergence or surfacing location [21]. Discrete search game problems can be formulated as follows: the sought-after object is hiding in one of the cells and the searcher sequentially inspects them.

One of the first solved search game problems is as follows: a plane is searching for an underwater submarine passing through a strait with varying width and of considerable length; the submarine cannot remain submerged throughout the entire crossing time; the task is to determine the optimal distribution of search efforts for the plane and the probability distribution of the submarine's submergence or surfacing location [21]. Discrete search game problems can be formulated as follows: the sought object is hidden in one of the cells, and the searcher sequentially examines them. When examining cell $i$ for a duration of $t_i$, the probability of detection, given that the target is in that cell, is equal to $p_i$. The payoff is the sum of m plus the duration of time spent examining the cells, $m + \sum_{k=1}^{m} t_{i_k}$, Where m is the number of cells viewed to detect the desired object. Bram [22] solved this problem for $p_i = 1$=1 and $t_i = 1$, Noitz [23] - for $0 < p_i \leq 1$ and $t_i \neq 0$. Johnson [24] - for $p_i = 1, t_i = 0$, provided that the seeker searcher receives additional information during each inspection regarding which of the two ordinal numbers of the cells is greater: the number of the cell in which the object is hidden, or the number of the cell that was inspected before the current inspection. Danskin considered the antagonistic game of distributing sequential search efforts among different areas, where false targets coexist with the desired object [25]. Ayzex [26] proposed a differential game of "Princess and Monster" search, which is formulated as follows. Monster P wants to catch Princess E. Both are located in a completely dark room (of any shape), but they know the boundaries of the room. The monster moves at a speed $v$, and the direction of movement can change instantly. The princess can move at any speed. The catch will occur if the distance between them becomes less than a given value $l$. The payment is the time required to catch the escaping E. In the simplified problem of "Princess and Monster", it is assumed that the pursuer R and the escaping E move along a circle. In the discrete variant of the simplified problem, P and E can occupy one of $n$ $(n \geq 3)$ points on the circle. Zelikin [27] solved the simplified problem under the condition that the speeds of the pursuer P and the escaping E are limited by the same value in absolute value, the positions at the initial moment follow a uniform distribution law, and the capture will occur if they end up in the same point $(l = 0)$. Alpern [28] considered the problem solved by Zelikin and showed that the absence of a limit on the speed of the escaping player does not affect the optimal strategies of the players.

Forman [29] conducts research on Zeilikin's strategy in the simplified problem of "Princess and Monster" where the cost for the pursuer P is the probability of not catching the evader E. In [30], Forman considers the problem of searching for "Princess and Monster" on a circle (with arbitrary initial positions and end time T) and in a certain area on a plane, assuming that the players are far from the boundary and cannot reach it in time T. The cost (for E) is the probability of being caught. Halpern considers a minimax problem that differs from the "Princess and Monster" problem in that E knows the trajectory of P [31]. For the case of a rectangular area, the "Princess and Monster" problem was solved by Gal [32]. Using this result, Fitzgerald obtained a solution for an arbitrary convex area, as well as for an area that is a finite union of convex areas [33].

Wilson [34] considers a wide class of differential search games of given duration, in which the players receive information about the initial position. He showed that such games have a solution in the class of mixed strategies and that a mixed strategy can be implemented by using a pure strategy whose choice depends on a randomly selected number from the unit interval.

This work is dedicated to the study of search problems where the target is mobile. Depending on the information available to the search participants, different approaches are proposed for finding a solution, i.e., determining optimal behavior.

## FUNDAMENTALS OF OBJECT SEARCH THEORY (BACKGROUND INFORMATION)

Subject of object search theory:

The subject of object search theory is the search for real objects in various environments. Search can be defined as the process of purposeful exploration of a specific area of space to detect an object located there. Detection refers to obtaining information about the location of an object by establishing direct energetic contact with it. Detection is carried out using detection means such as optical, radar, hydroacoustic and other devices.

One way to study the search process is to build and analyze mathematical models that reflect the objective laws of search and allow us to establish causal relationships between the conditions of search performance and its results. The search process involves two sides: the search object and the observer, who can be both individual and group. The search object is various objects located in different environments, such as aircraft, various objects on the surface of the Earth, ships, and vessels, etc.

The search object has two characteristic features:

Its properties differ from the properties of the environment in which the search is carried out.
Information about the location of the object before the start of the search and during its execution is usually uncertain.

It is this uncertainty that causes search actions, the essence of which is to obtain information about the location of the object. The contrast of the search object against the background of the environment creates the possibility of its detection.

Search objects can be characterized by the presence or absence of radiation. Therefore, the operation of detection tools is based either on the detection of a signal reflected from the search object or on the reception of the object's own radiation. The search process largely depends on the properties of the detection object, as well as on the parameters of the detection equipment and the characteristics of the surrounding environment. All these issues form the physical basis of the theory of search.

During the search process, the use of detection tools is combined with the active maneuvering of the observer who carries these tools. Therefore, the study of the patterns of mutual movement of the observer and the search object becomes especially important. These patterns constitute an integral part of the theory of object search - the kinematics of search.

An important place in the theory of object search is occupied by the justification and methods of calculating the indicators of the success of the search - criteria of its effectiveness. The ultimate goal of the theory of search is to choose the optimal methods for performing search actions in a specific situation and under the conditions of the search - the so-called search situation. The choice of the optimal search method is based on an analysis of the mathematical model of the corresponding search situation and reduces to establishing control parameters of the search that ensure the solution of the search task in the shortest or specified time with minimal search efforts.

Mathematical Models for Object Search

Constructing a mathematical model requires identifying all the essential factors and conditions that determine both the state and the development of the search process, as well as the possible control of this process. These factors and conditions are variables and are called elements of the model. The variables that can be changed are called controllable, while those that cannot be changed are called uncontrollable. Depending on the nature of the search process, its mathematical model may contain only uncontrollable variables, or both controllable and uncontrollable ones. Mathematical models of the first type are called descriptive, while models of the second type are normative.

In a descriptive model, there is no observer deciding about the search, nor is there a search object deciding to evade.

A normative model is characterized by the presence of at least one of the parties making a decision. Depending on the amount of information about the search situation and the regularities underlying it, such models can be classified into one of the following levels: deterministic, stochastic, and uncertain.

At the deterministic level, a normative model is constructed when the outcome of the search situation is subject to regularities and the factors influencing this outcome can be accurately measured or estimated, while random factors are either absent or can be neglected. In this case, it is quite difficult to collect and process data, except for simple situations that fully characterize the search conditions. In addition, purely mathematical difficulties arise in constructing and analyzing such a model. The task of choosing the optimal search method in the conditions of a normative model of the deterministic level is reduced to maximizing or minimizing the efficiency criterion.

At the stochastic level, the normative model, in accordance with probabilistic regularities, is represented as a random process, the course and outcome of which are described by certain characteristics of random variables. Construction of a model at this level is possible if there is sufficient factual material to estimate the necessary probability distributions. In constructing a model at the stochastic level, the method of statistical trials is widely used in addition to the classical apparatus of probability theory, and the principle of optimality is based on the maximization of the mathematical expectation of the efficiency criterion. Thus, the task is practically transferred to the deterministic level.

The indeterminate level is characterized by such a volume of information at which only a set of possible search situations is known, but without any a priori information about the probability of each of them. Usually, such a volume of information is characteristic of a conflict situation in which the object of the search and the observer pursue directly opposing goals, choosing a certain way of action to achieve them.

Building a model and choosing the optimal way at the indeterminate level encounters some difficulties, since the principles of optimality in this case may not be entirely clear. Establishing such principles of optimality and finding solutions to such problems constitute the content of game theory.

Game theory deals with the study of mathematical models of decision-making in conditions of conflict. To construct a formal mathematical model of decision-making in conditions of conflict, it is necessary to mathematically describe all possible actions of the participants in the conflict and the results of these actions. The results of the players' actions are evaluated using a numerical function called the payoff function.

## TASK FORMULATION

Let's move on to the description of the search process, which is the focus of the main part of this work. A ship-interceptor equipped with a hydro locator has detected the periscope of a submarine, which immediately disappeared in an unknown direction. A hydro locator is a means of sound detection of underwater objects using acoustic radiation. It consists of a transceiver that sends sound pulses in the required direction and receives reflected pulses if the transmission, encountering any object on its path, is reflected from it. After the initial detection of the submarine, the task of the ship-interceptor is to catch the submarine in the shortest possible time. It is assumed that although the ship does not know the exact speed of the submarine, it knows a discrete set of speeds, one of which is the actual speed of the submarine. The formulated problem is a problem of secondary search for a moving object. The ship-interceptor will be referred to as the pursuer and the submarine as the evader, denoted as P and E, respectively.

The work considers cases of continuous search, when the resistance of the boat and the ship is not considered, game problems of search, and the case of searching for n submarines. Thus, the goals of my research are the mathematical formalization of the process of search and interception of moving objects under various information conditions; the development of a procedure for finding the optimal solution; the implementation of the algorithm using the MAPLE 17 software package.

## ONE PURSUER AND ONE EVADER SCENARIO

Algorithm for finding the guaranteed capture time.

Let us present an algorithm for finding the completion time of the search under conditions when the pursuer does not know the speed of the evader with certainty. To do this, let us first show the strategy of the pursuer's behavior. Assume that the speed of the pursuer is so much greater than the speed of the evading submarine that the completion of the search is guaranteed. At the initial moment $t_0$ of detection, the pursuer P accurately determines the location of the underwater submarine. Thus, he knows the distance $D_0$ between him and the evader. To find the guaranteed completion time of the search, let us build a normative model of a deterministic level. Let us introduce a polar coordinate system (ρ, φ, O) such that the pole O is located at the point of detection of the underwater submarine, and the polar axis ρ passes through the point where the intercepting ship is located. The pursuer does not know the speed v of the evader with certainty, but it is known that it is chosen from the discrete set $V^E = \{v_1, \dots, v_n\}$. The maximum possible speed of the pursuer ship is denoted by $v^P$. Then, the dynamics of the evader E are described by equations:

$$\dot{\rho}^E = v$$

$$\dot{\varphi}^E = 0$$

The dynamics of the pursuer is described by the equations.

$$\dot{\rho}^P = \alpha, |\alpha| \leq v_\rho$$

$$\dot{\varphi}^P = \beta, |\beta| \leq v_\varphi$$

$$v^P = \sqrt{(v_\rho)^2 + (v_\varphi)^2}$$

Since the speed of the fleeing vessel is not known with certainty, the pursuer makes the assumption that E has a speed of $v_1 \in V^E$. To capture the submarine at time $t_0$, the pursuer begins moving towards point O with a speed of $v^P$ and continues until time $t_1$, at which point both players are at the same distance from point O, i.e., the equation.

$$\rho_1^P = \rho_1^E$$

And

$$\int_{t_0}^{t_1} v_1 dt + v^P(t_1 - t_0) = D_0$$

If the encounter did not occur, then now $t_1$, the pursuer, choosing a direction of circumnavigation, continues to move around point O in such a way as to constantly remain at the same distance from point O as the fleeing ship. Let's find the trajectory of motion corresponding to this behavior strategy. We will consider the direction of circumnavigation coinciding with the positive direction of the polar angle. The speed of the interceptor ship can be decomposed into two components: radial $v_\rho$ and tangential $v_\varphi$. The radial component is the speed at which the ship moves away from the pole, i.e.

$$v_\rho = \dot{\rho}$$

The tangential component is the linear speed of rotation relative to the pole, i.e.

$$v_\varphi = \rho\dot{\varphi}$$

In order for the encounter to occur, the pursuer moves at maximum speed, keeping the radial component of the velocity equal to the speed of the fleeing vessel. Then, to find the trajectory of the pursuer, it is necessary to solve the system of differential equations:

$$\dot{\rho} = v_1$$

$$\dot{\varphi}^2 \rho^2 = (v^P)^2 - (v_1)^2$$

The initial conditions for this system are.

$$\varphi(t^*) = 0$$

$$\rho(t_1) = v_1 t_1$$

Solving it, we find:

$$\varphi(t) = \frac{\sqrt{(v^P)^2 - (v_1)^2}}{v_1} \ln \frac{v_1 t}{v_1 t_1}$$

$$\rho(t) = v_1 t$$

Then the search time can be expressed as a function of the polar angle:

$$t(\varphi) = t_1 \exp\left(\frac{v_1 \varphi}{\sqrt{(v^P)^2 - (v_1)^2}}\right)$$

Thus, the trajectory consists of straight-line segments and logarithmic spiral segments. By adhering to this behavior strategy, the pursuer will detect the submarine within a time not exceeding one spiral turn. Then, if the ship, having bypassed the spiral turn, does not find the submarine, it means that the initial assumption about the speed of the evader was incorrect. Therefore, it is necessary to choose the next speed $v_2 \in V^E$ and assume that it is the actual speed. The evader has covered a distance of $\rho_E(t_2) = v_2 t_2$, during time t_2, while the pursuer has covered $\rho_P(t_2) = v_1 t_2$. There are two cases. If $\rho_P(t_2) > \rho_E(t_2)$, then the distance between the players will be equal to $D_2 = \rho_P(t_2) - \rho_E(t_2)$ and to find the time t_3, the equation must be solved:

$$\int_{t_2}^{t_3} v_2 dt + v^P(t_3 - t_2) = D_2$$

If $\rho_P(t_2) < \rho_E(t_2)$, then the distance between the players will be equal to $D_2 = \rho_E(t_2) - \rho_P(t_2)$, and to find the moment in time $t_3$, we need to solve the equation:

$$v^P(t_3 - t_2) - \int_{t_2}^{t_3} v_2 dt = D_2$$

$V^E$. This algorithm for computing the guaranteed capture time is implemented using the software package MAPLE 17.

In our task, the number of speeds n is finite and known in advance, and each speed needs to be checked for validity. It is assumed that when creating the schedule, the processing can start with any of the speeds. The duration of checking each speed will depend on the created schedule. Using the algorithm for finding the guaranteed search time outlined above, we will construct a matrix of times $T = (t_{ij})$, where $t_{ij}$ represents the duration of checking the speed when speed $v_i$ precedes speed $v_j$. Then the time taken by the pursuer to check all speeds, i.e., Guaranteed search time, depends on the order. Denote by $F_{max} = F_{[n]} = \sum_{i=1}^{n} t_{[i-1],[i]}$ maximum test duration. The task is to check each of the n speeds once and only once, and the order should be such as to minimize the maximum duration of the passage. It is necessary to find such a matrix $X$ of order $n$ with elements.

$$x_{ik} = 0 \text{ или } 1, i = 1, \ldots, n, k = 1, \ldots, n$$

$$\sum_{i=1}^{n} x_{ik} = 1, k = 1, \ldots, n$$

$$\sum_{k=1}^{n} x_{ik} = 1, i = 1, \ldots, n$$

to the sum $\sum_{i=1}^{n}\sum_{k=1}^{n} x_{ik}t_{ik}$ is minimized. Two methods are used to obtain optimal solutions with high accuracy for small numbers of speeds and approximate solutions for large ones. The first algorithm is called the branch and bound method, which involves sequentially transforming matrices into one of three standard forms by a certain procedure. First, an initial matrix is constructed based on the given problem. Then, the matrix is processed according to a specific scheme to obtain simpler variations, also represented in the form of matrices. The standard procedures are repeatedly applied to each of these variations until a final solution to the problem is obtained. Thus, the analysis of each matrix leads to one of the three possibilities:

1. Obtaining a solution directly from the original matrix when the problem is simple enough.
2. Excluding a matrix from further consideration when it can be shown that it does not lead to a solution of the problem.
3. Branching, which involves reducing the problem to considering two less complex variants of the problem.

The solution to the problem involves checking all speeds, given by a permutation of indices 1..., n.

$$[1], [2], \ldots, [n]$$

The obtained solution is a sum of n terms, each of which is determined by an element of the matrix T according to the adopted order:

$$t_{[1],[2]} + t_{[2],[3]}, + \cdots + t_{[n-1],[n]}$$

The optimal solution is the permutation that minimizes this sum. At each step of the algorithm, the problem involves n speeds, of which k can be determined, and the remaining n-k must be chosen optimally. For all selections, a value Y must be assigned as the lower bound for all possible solutions to the problem, including the optimal solution. There are trivial lower bounds, such as the minimum element of matrix T or the sum of the minimum elements of its n rows. The subtlety of the algorithm lies in constructing this lower bound, striving to make it as large as possible.

Thus, the matrix is characterized by the remaining number of unknown steps for checking, $n - k$, and the lower bound Y of the solution. Moreover, it can be assumed that for the remaining set of steps, at least one solution to the problem is known (for example, the permutation 1..., n is a solution), and let Z be the best of them. Now the matrix undergoes further changes depending on the following possibilities:

If $n - k = 2$, then there are no more than two steps left and the solution is found immediately. If its value is less than Z, then Z is set equal to this new value and is considered the best of the known solutions.

If Y is greater than or equal to Z, the matrix is excluded because the checks presented in it do not lead to better solutions than what is already known.

If none of the above situations apply, then two matrices are created instead of the original one. The branching of the original verification occurs in two directions, and each direction corresponds to its own matrix:

In one of them, the transition from i to j is chosen, as a result of which the lower bound of the solutions may increase.

In the other, the transition from i to j is prohibited (the element $t_{ij}$ is set equal to $\infty$), because of which the lower bound of the solutions will undoubtedly increase.

Thus, the resulting matrices are characterized by an increasing lower bound and (or) a larger number of established steps. In addition, for each subsequent matrix, the number of checks is less than for the previous one, and eventually, a state is reached where the permutation is fully defined.

The situations where the solution is obtained immediately, or the matrix is excluded are obvious. The essence of branching is the concepts of reduction and selection. The reduction aims to obtain at least one zero in each row and column of the original matrix T. Since each solution to the problem includes one and only one element from each row or column of matrix T, subtracting or adding a constant to each element of its column or row in the same degree changes all solutions and does not lead to a displacement of the optimum.

Subtract a constant h from each element of a row or column of matrix T. Let the resulting matrix be $T'$. Then the optimal solution found from $T'$ is also optimal for T, i.e., both matrices have the same permutation that minimizes time. We can choose $Y' = h$ as the lower bound for solutions obtained from $T'$. Subtraction can continue until each column or row contains at least one zero (i.e., the minimum element in each row or column is zero). The sum of all reduction constants determines the lower bound Y for the original problem. The matrix T is reduced if it cannot be further reduced. In this case, finding route options is associated with studying a particular transition, say from i to j. As a result, instead of the original matrix, we consider two matrices:

1. Matrix $T_{ij}$, which is associated with finding the best of all solutions given by matrix T and including the order (i, j).
2. Matrix $T_{n(ij)}$, which is associated with choosing the best of all solutions not including the order (i, j).

After fixing the transition from i to j, we need to exclude transitions from i to other speeds except j, and transitions to j from other speeds except i, by setting all elements of row i and column j, except $t_{ij}$, to infinity. We also need to prohibit the order (j, i) in the future by setting $t_{ij} = \infty$. This is because checking all speeds during a single pass cannot include both (i, j) and (j, i) simultaneously. Since these prohibitions may lead to the elimination of some zeros in matrix T, further reduction of T and obtaining a new, larger lower bound for solutions associated with matrix $T_{ij}$ is not excluded.

In the matrix $T_{n(ij)}$, it is prohibited to transition from i to j, i.e., $t_{ji}$ is set to infinity. In this case, the possibility of further reducing the matrix and the resulting increase in the lower bound for solutions obtained from $T_{n(ij)}$ is not excluded. The choice of (i, j) should be such as to maximize the lower bound for $T_{n(ij)}$, which may allow for the elimination of trajectories without further branching. To achieve this, all possible pairs (i, j) in the matrix $T_{n(ij)}$ are examined, and the choice is made in such a way that the sum of two consecutive reducing constants is maximal. Obviously, transitions (i, j) corresponding to zero elements of matrix T should be prohibited first, since the choice with nonzero elements does not contribute to further reducing $T_{n(ij)}$.

The second way to order the enumeration of velocities is the method of dynamic programming. Without loss of generality, choose a certain velocity $v_0$ as the initial one. After that, divide the set of all velocities into four non-intersecting subsets:

In the matrix $T_{n(ij)}$ the transition from i to j is forbidden, i.e. $t_{ji}=\infty$ is assumed. In this case, there is also the possibility of further reducing the matrix and the resulting increase in the lower bound for solutions obtained from $T_{n(ij)}$. The choice of (i, j) should be such as to maximize the lower bound for $T_{n(ij)}$, which may allow the exclusion of a number of trajectories without further branching. To achieve this, all possible pairs (i, j) in the matrix $T_{n(ij)}$ are examined, and the choice is made in such a way that the sum of two consecutive leading constants is maximized. It is obvious that transitions (i, j) that correspond to zero elements of the matrix T should be prohibited in the first place since the choice of non-zero elements do not contribute to further reduction of $T_{n(ij)}$.

The second method for ordering the enumeration of speeds is the dynamic programming approach. Without loss of generality, we choose some speed $v_0$ as the initial speed. After that, we divide all the set of velocity into four disjoint subsets:

$\{v_0\}$ - the set consisting only of the initial speed.

$\{v_i\}$ - the set consisting only of one non-initial speed.

$\{V_k\}$ - the set consisting of k speeds, except for $v_0$ and $v_i$

$\{V_{n-k-2}\}$ - the set consisting of the remaining n-k-2 speeds.

Let us assume that the optimal order of checking speeds is known, starting with speed $v_0$. Then we can choose speed $v_0$ and a subset $\{V_k\}$ consisting of k speeds, in such a way that this optimal permutation begins with $\{v_0\}$ and includes the set $\{V_{n-k-2}\}$, then $\{v_i\}$, after which it checks the set $\{V_k\}$.

Now let us consider only the part of the permutation that lies between $\{v_i\}$ and $\{v_0\}$ with an intermediate check of $\{V_k\}$. It can be noted that the minimum time for this segment is known. If this were not the case, then without changing the part of the permutation up to speed $v_i$, we could find the best guaranteed time for completing its check and, therefore, the minimum time for the whole. However, this is impossible, since it contradicts the initial assumption that the optimal permutation is known.

Let f $(v_i; \{V_k\})$ be the time for checking the best permutation from $v_i$ to $v_0$, including the set $\{V_k\}$. Note that when k=0,

$$f(v_i; \{\emptyset\}) = s_{i0}$$

If T is an element of the matrix T, and k=n-1 and $v_i$ coincides with the start of the movement, then f($v_0;\{V_{n-1}\}$) is the time of the optimal permutation of the original problem. The idea of dynamic programming is to increment k step by step, starting from k=0. Starting from $v_0$, the permutation is traversed in reverse order to find the optimal solution.

For the problem under consideration, the main functional equation of dynamic programming is given by:

$$f(v_i; \{V_k\}) = \min_{v_j \in \{V_k\}}[s_{ij} + f(v_j; \{V_j\} - \{v_j\})]$$

This equation shows that to find the best permutation starting from $v_i$ and ending with $v_0$, with k intermediate velocities, one needs to choose the best among k permutations, starting from the transition from $v_i$ to one of the k velocities and then moving the fastest way to $v_0$ with intermediate visits to k-1 others. Each of these k options, in turn, represents the fastest of k-1 permutations according to the equation mentioned earlier. Eventually, a point is reached where the right-hand side of the equation simply represents an element of T.

The solution to the problem for five velocities will be considered as an example, with the fifth velocity taken as the starting point. Then, f ($v_5; \{v_1, v_2, v_3, v_4\}$) represents the shortest time for the best permutation, and any sequence of checking velocities that leads to such time is optimal. At step 0, the solution is sought for five options with k=0:

$$f(v_1; \{\emptyset\}) = t_{16}$$

$$f(v_2; \{\emptyset\}) = t_{26}$$

$$f(v_3; \{\emptyset\}) = t_{36}$$

$$f(v_4; \{\emptyset\}) = t_{46}$$

At the first step, solutions for k=1 are expressed in terms of known solutions for k=0:

$$f(v_1; \{v_2\}) = t_{12} + f(v_2; \{\emptyset\})$$

$$f(v_1; \{v_3\}) = t_{13} + f(v_3; \{\emptyset\})$$

$$f(v_1; \{v_4\}) = t_{14} + f(v_4; \{\emptyset\})$$

$$f(v_2; \{v_1\}) = t_{21} + f(v_1; \{\emptyset\})$$

$$f(v_2; \{v_3\}) = t_{23} + f(v_3; \{\emptyset\})$$

....

$$f(v_4; \{v_3\}) = t_{43} + f(v_3; \{\emptyset\})$$

At the second step, solutions for k=2 are expressed in terms of known solutions for k=1:

$$f(v_1; \{v_2, v_3\}) = min[t_{12} + f(v_2; \{v_3\}), t_{13} + f(v_3; \{v_2\})]$$

$$f(v_1; \{v_2, v_4\}) = min[t_{12} + f(v_2; \{v_4\}), t_{14} + f(v_4; \{v_2\})]$$

$$f(v_1; \{v_2, v_5\}) = min[t_{12} + f(v_2; \{v_5\}), t_{15} + f(v_5; \{v_2\})]$$

$$f(v_1; \{v_3, v_4\}) = min[t_{13} + f(v_3; \{v_4\}), t_{14} + f(v_4; \{v_3\})]$$

$$f(v_1; \{v_3, v_5\}) = min[t_{13} + f(v_3; \{v_5\}), t_{15} + f(v_5; \{v_3\})]$$

$$f(v_1; \{v_4, v_5\}) = min[t_{14} + f(v_4; \{v_5\}), t_{15} + f(v_5; \{v_4\})]$$

$$f(v_2; \{v_1, v_4\}) = min[t_{21} + f(v_1; \{v_4\}), t_{24} + f(v_4; \{v_1\})]$$

....

$$f(v_4; \{v_2, v_3\}) = min[t_{42} + f(v_2; \{v_3\}), t_{43} + f(v_3; \{v_2\})]$$

We proceed to the third step, using each of the solutions of the second step.

$$f(v_1; \{v_2, v_3, v_4\}) = min[t_{12} + f(v_2; \{v_3, v_4\}), t_{13} + f(v_3; \{v_2, v_4\}),$$

$$t_{14} + f(v_4; \{v_2, v_3\})]$$

$$f(v_2; \{v_1, v_3, v_4\}) = min[t_{21} + f(v_1; \{v_3, v_4\}), t_{23} + f(v_3; \{v_1, v_4\}),$$

$$t_{24} + f(v_4; \{v_1, v_3\})]$$

$$f(v_3; \{v_1, v_2, v_4\}) = min[t_{31} + f(v_1; \{v_2, v_4\}), t_{32} + f(v_2; \{v_1, v_4\}),$$

$$t_{34} + f(v_4; \{v_1, v_2\})]$$

$$f(v_4; \{v_1, v_2, v_3\}) = min[t_{41} + f(v_1; \{v_2, v_3\}), t_{42} + f(v_2; \{v_1, v_3\}),$$

$$t_{43} + f(v_3; \{v_1, v_2\})]$$

At the fourth step, the solution of the original problem is obtained.

$$f(v_5; \{v_1, v_2, v_3, v_4\}) = min[t_{51} + f(v_1; \{v_2, v_3, v_4\}), t_{52} + f(v_2; \{v_1, v_3, v_4\}),$$

$$t_{53} + f(v_3; \{v_1, v_2, v_4\}), t_{54} + f(v_4; \{v_1, v_2, v_3\})]$$

Exists $\frac{(n-1)!}{k!(n-k-2)!}$ variations. For k≥1 there are k choices.

The number of comparisons to be made between them. The total number of computations at all stages will be equal to this number.

$$2 \sum_{k=1}^{n-1} \frac{k(n-1)!}{k!(n-k-2)!} + (n-1) < n^2 2^n$$

As an example, let us consider solving the problem for six speeds $V^E = \{10,20,30,40,50,60\}$. The initial matrix T is obtained by applying the algorithm for computing the guaranteed search time, implemented using the Maple software package.

$$\begin{matrix} 0.13 & 0.42 & 1.13 & 3.05 & 9.13 & 34.1 \\ 0.25 & 0.24 & 1.74 & 4.73 & 14.23 & 53.27 \\ 0.54 & 1.29 & 0.44 & 7.6 & 22.94 & 86.04 \end{matrix}$$

$$\begin{matrix} 1.23 & 2.84 & 6 & 0.89 & 39.15 & 147.2 \\ 3.14 & 7.04 & 14.7 & 31.4 & 2 & 277.2 \\ 9.62 & 21.16 & 43.89 & 3.13 & 217.76 & 5.6 \end{matrix}$$

Theoretical game model of search and interception.

To reduce the guaranteed interception time, it is advisable for the pursuer to order the search of escape speeds. However, if the escapee becomes aware of this, they can move at a speed that the pursuer intends to check last, which would allow the escapee to maximize their search time. Thus, the search problem can be considered a game problem in the conditions of opposition.

The system G = (X, Y, K), where X and Y are non-empty sets and the function K: X × Y → R1 is called an antagonistic game in normal form. Elements x ∈ X and y ∈ Y are called player 1 and player 2 strategies, respectively, in game G. The Cartesian product elements (i.e., strategy pairs (x, y), where x ∈ X and y ∈ Y) are called situations, and the function K is the function of player 1's gain. Player 2's gain in an antagonistic game in situation (x, y) is assumed to be [-K (x, y)], so the function K is also called the game's gain function, and game G is a zero-sum game.

Let's define the game for the search problem under consideration. Let the escapee choose any speed from the set $V^E = \{v_1, ..., v_n\}$ and any direction from the set $\alpha = \{\alpha_1, ..., \alpha_n\}$. Then, the set of pure strategies for the escapee (player 1) will be the set of combinations of possible velocities $v_i$ of their movement and movement directions α, and the set of pure strategies for the pursuer will be the set of all possible permutations of the escapee's velocities. The gain will be the time it takes to catch the escapee, which is found using the algorithm described above. The game G is interpreted as follows: players independently and simultaneously choose strategies x ∈ X and y ∈ Y. After that, player 1 receives a gain equal to K(x, y), and player 2 receives a gain equal to (-K(x, y)). Antagonistic games in which both players have finite sets of strategies are called matrix games.

Let player 1 in a matrix game have a total of m strategies. We establish a one-to-one correspondence between the set X of strategies and the set M = {1, 2, ..., m}. Similarly, if player 2 has n strategies, we can establish a one-to-one correspondence between the sets N = {1, 2, ..., n} and Y. Then, the game G is fully determined by the matrix A = $\{\alpha_{ij}\}$, where $\alpha_{ij} = K(x_i, y_j), (i,j) \in M \times N, (x_i, y_j) \in X \times Y, i \in M, j \in N$. In this case, the game G is played as follows: player 1 chooses a row i

∈ M, and player 2 (simultaneously with player 1) chooses a column j ∈ N. After that, player 1 receives a payoff of $\alpha_{ij}$, and player 2 receives ($-\alpha_{ij}$).

Each player aims to maximize their own winnings by choosing a strategy. However, for player 1, their winnings are determined by the function K(x,y), while for the second player it is (-K(x,y)), i.e. the players' goals are directly opposite. It should be noted that the winnings of player 1 (2) are determined in the situations (x,y)∈X×Y that arise during the game. However, each situation, and therefore the winnings of a player, depend not only on their own choice, but also on what strategy their opponent will choose. Therefore, in seeking to obtain the maximum winnings possible, each player must take into account the behavior of their opponent.

In game theory, it is assumed that both players act rationally, i.e., strive to achieve maximum winnings, assuming that their opponent acts in the best possible way for themselves. Let player 1 choose a strategy x. Then in the worst case, they will win $\min_{y} K(x,y)$. Therefore, player 1 can always guarantee themselves a win of $\max_{x} \min_{y} K(x,y)$. If we abandon the assumption of the attainability of the extremum, then player 1 can always obtain winnings that are arbitrarily close to this value.

$$\underline{v} = \sup_{x \in X} \inf_{y \in Y} K(x,y)$$

which is called the lower value of the game. If the external extremum is reached, then the value ⎯v is also called the maximin, the principle of constructing the strategy x, based on maximizing the minimum payoff, is called the maximin principle, and the strategy x chosen in accordance with this principle is the maximin strategy of player 1.

For player 2, similar reasoning can be applied. Suppose they choose strategy y. Then in the worst case, they will lose $\max_{x} K(x,y)$. Therefore, the second player can always guarantee a loss of $\min_{y} \max_{x} K(x,y)$. The number... (the text appears to be cut off here, so I am unable to translate the complete sentence).

$$\overline{v} = \inf_{y \in Y} \sup_{x \in X} K(x,y)$$

The upper value of the game G is called the maximum-minimum, and in the case of achieving an external extremum, it is called the minimax. The principle of constructing the strategy y, based on minimizing the maximum losses, is called the minimax principle, and the strategy y chosen in accordance with this principle is the minimax strategy of player 2. It should be emphasized that the existence of a minimax (maximin) strategy is determined by the achievability of an external extremum. In the matrix game G, the extremums are achieved, and the lower and upper values of the game are respectively equal.

$$\overline{v} = \min_{1 \leq j \leq n} \max_{1 \leq i \leq m} \alpha_{ij}$$

$$\underline{v} = \max_{1 \leq i \leq m} \min_{1 \leq j \leq n} \alpha_{ij}$$

The minimax and maximin for the game G can be found as follows:

$$\begin{bmatrix} \alpha_{11} & \cdots & \alpha_{1n} \\ \vdots & \ddots & \vdots \\ \alpha_{m1} & \cdots & \alpha_{mn} \end{bmatrix} \left.\begin{matrix} \min_{j} \alpha_{1j} \\ \cdots \\ \min_{j} \alpha_{mj} \end{matrix}\right\} \max_{i} \min_{j} \alpha_{ij} = \underline{v}$$

$$\max_{i} \alpha_{i1} \quad \ldots \quad \max_{i} \alpha_{in} \} \min_{j} \max_{i} \alpha_{ij} = \overline{v}$$

Let's consider the question of optimal behavior of players in an antagonistic game. It is natural to consider a situation $(x^*, y^*) \in X \times Y$ in game G=(X,Y,K) optimal if neither player has an incentive to deviate from it. Such a situation $(x^*, y^*)$ is called an equilibrium, and the optimality principle based on constructing an equilibrium situation is called the principle of equilibrium. For antagonistic games, the principle of equilibrium is equivalent to the principles of minimax and maximin. In an antagonistic game G=(X, Y,K), a situation $(x^*, y^*)$ is called an equilibrium or a saddle point if

$$K(x, y^*) \leq K(x^*, y^*)$$

$$K(x^*, y) \geq K(x^*, y^*)$$

or all x∈X,y∈Y.

For the matrix game G, we are talking about the saddle points of the payoff matrix A, i.e., points $(i^*, j^*)$ such that for all i∈M, j∈N the inequalities.

$$\alpha_{ij^*} \leq \alpha_{i^*j^*} \leq \alpha_{i^*j}$$

Theorem. Let $(x_1^*, y_1^*)$ and $(x_2^*, y_2^*)$ be two arbitrary equilibrium situations in the antagonistic game G. Then

$$K(x_1^*, y_1^*) = K(x_2^*, y_2^*); K(x_1^*, y_2^*) = K(x_2^*, y_1^*)$$

$(x_1^*, y_2^*) \in Z(G), (x_2^*, y_1^*) \in Z(G)$ where $Z(G)$ the set of all equilibrium situations.

Let $(x^*, y^*)$ be an equilibrium situation in game G. The number $v = K(x^*, y^*)$ is called the value of game G.

Now we establish a connection between the principle of equilibrium and the principles of minimax in an antagonistic game.

Theorem. In order for there to exist an equilibrium situation in game $G = (X, Y, K)$, it is necessary and sufficient that the minimax and maximin $min_y sup_x K(x, y), max_x inf_y K(x, y)$ exist, and the equality is satisfied:

$$\overline{v} = max_x \, inf_y \, K(x, y) = min_y \, sup_x \, K(x, y) = \overline{v}$$

If there exists a situation of equilibrium in a matrix game, then the minimax is equal to the maximin, and according to the definition of equilibrium situation, each player can communicate their optimal (maximin) strategy to their opponent, and from this neither player can gain any additional advantage. Now, suppose that in game G there is no situation of equilibrium. Then, we have

$$min_j \, max_i \, \alpha_{ij} - max_i \, min_j \, \alpha_{ij} > 0$$

If there is an equilibrium situation in a matrix game, then the minimax is equal to the maximin, and according to the definition of the equilibrium situation, each player can inform their optimal (maximin) strategy to the opponent, and neither player can gain additional advantage from this. Now suppose there is no equilibrium situation in game G. In this case, the maximin and minimax strategies are not optimal. Moreover, it may not be beneficial for the players to adhere to them, as they may obtain a greater gain. However, informing the opponent about the choice of strategy can lead to even greater losses than in the case of the maximin or minimax strategy.

In this case, it is reasonable for players to act randomly, which provides the greatest secrecy in choosing a strategy. The result of the choice cannot become known to the opponent, as the player themselves do not know it until the random mechanism is implemented. A random variable, whose values are the player's strategies, is called their mixed strategy. Since a random variable is characterized by its distribution, we will identify a mixed strategy x of player 1 in a game with an m-dimensional vector.

$$x = (\xi_1, \ldots, \xi_m) \in R^m, \sum_{i=1}^{m} \xi_i = 1, \xi_i \geq 0, i = 1, \ldots, m$$

Similarly, player 2's mixed strategy y is the n-dimensional vector.

$$y = (\eta_1, \ldots, \eta_n) \in R^n, \sum_{j=1}^{n} \eta_j = 1, \eta_j \geq 0, j = 1, \ldots, n$$

In this case, $\xi_i \geq 0$ and $\eta_j \geq 0$ are the probabilities of choosing pure strategies $i \in M$, $j \in N$ respectively when players use mixed strategies x and y. Let X and Y denote the sets of mixed strategies for the first and second players respectively. Let $x=(\xi_1,\ldots, \xi_m) \in X$ be a mixed strategy. The set of mixed strategies for a player is an extension of their pure strategy space. A pair (x, y) of mixed strategies for players in matrix game G is called a situation in mixed strategies.

Let's define the payoff of player 1 in the situation (x, y) in mixed strategies for the matrix game G as the mathematical expectation of their payoff given that players use mixed strategies x and y respectively. The players choose their strategies independently of each other, therefore the expected payoff K (x, y) in the situation (x, y) in mixed strategies $x=(\xi_1,\ldots, \xi_m)$ and $y=(\eta_1,\ldots, \eta_n)$ is equal to:

$$K(x,y) = \sum_{i=1}^{m}\sum_{j=1}^{n} \alpha_{ij}\xi_i\eta_j$$

The situation $(x^*, y^*)$ is called an equilibrium situation if

$$K(x, y^*) \leq K(x^*, y^*)$$

$$K(x^*, y) \geq K(x^*, y^*)$$

for all $x \in X, y \in Y$.

Theorem. Every matrix game has a situation of equilibrium in mixed strategies.

A common way to solve a matrix game is by reducing it to a linear programming problem. However, difficulties arise when solving matrix games of large dimensions. Therefore, the iterative Brown-Robinson method is often used to find a solution. The idea of the method is to repeatedly play a fictitious game with a given payoff matrix. One repetition of the game is called a round. Let $A = \{\alpha_{ij}\}$ be an (m x n)-matrix game. In the first round, both players choose their pure strategies completely randomly. In the k-th round, each player chooses the pure strategy that maximizes their expected payoff against the observed empirical probability distribution of the opponent's moves in the previous (k-1) rounds.

So, suppose that in the first k rounds, player 1 used the i-th strategy $\xi_i^k$ times, and player 2 used the j-th strategy $\eta_j^k$ times. Then in the (k+1)-th round, player 1 will use the $i_{k+1}$-th strategy, and player 2 will use their $j_{k+1}$ strategy, where:

$$\overline{v}^k = \max_i \sum_j \alpha_{ij}\eta_j^k = \sum_j \alpha_{i_{k+1}j}\eta_j^k$$

$$\underline{v}^k = \min_j \sum_i \alpha_{ij}\xi_j^k = \sum_j \alpha_{ij_{k+1}}\xi_j^k$$

Let v be the value of the matrix game G. Consider the relations

$$\overline{v}^k/k = \max_i \sum_j \alpha_{ij}\eta_j^k/k = \sum_j \alpha_{i_{k+1}j}\eta_j^k/k$$

$$\overline{v}^k/k = \min_j \sum_i \alpha_{ij}\xi_j^k/k = \sum_j \alpha_{ij_{k+1}} \xi_j^k/k$$

Vectors $x^k = (\frac{\xi_1^k}{k}, \ldots, \frac{\xi_m^k}{k})$ и $y^k = (\frac{\eta_1^k}{k}, \ldots, \frac{\eta_n^k}{k})$ are mixed strategies of players 1 and 2, respectively, so by the definition of the value of the game we have

$$\max_k \underline{v}^k/k \leq v \leq \min_k \overline{v}^k/k$$

Thus, an iterative process is obtained, which allows finding an approximate solution of the matrix game, and the degree of approximation to the true value of the game is determined by the length of the interval. The convergence of the algorithm is guaranteed by a theorem.

$$\lim_{k\to\infty}(\min_k \overline{v}^k/k) = \lim_{k\to\infty}(\max_k \underline{v}^k/k) = v$$

Consider some examples of solving pursuit games.

**Example 1**. Let the initial distance between the pursuer and the evader be 200 km. The evader chooses a velocity from the set $V^E$= {8, 56, 78} and a direction from the set α= {23, 137, 182}. The maximum speed of the pursuer is $V^P$=100 km/h. Then the set of strategies for the evader is:

$$(\alpha_1, v_1), (\alpha_1, v_2), (\alpha_1, v_3), (\alpha_2, v_1), (\alpha_2, v_2), (\alpha_2, v_3), (\alpha_3, v_1), (\alpha_3, v_2),$$

set of pursuer strategies:

$$(v_1, v_2, v_3), (v_1, v_3, v_2), (v_2, v_1, v_3), (v_2, v_3, v_1), (v_3, v_1, v_2), (v_3, v_2, v_1)$$

The resulting game matrix looks like this:

|  |  |  |  |  |  |
|---|---|---|---|---|---|
| 1,9 | 1,9 | 134 768423,4 | 4817,7 | 336724,9 |
| 8,4 | 48345 | 1,7 | 1,7 | 21184,8 | 4236,3 |
| 1478 | 21 | 1478 | 295,6 | 1,9 | 1,85 |
| 2,2 | 2,2 | 156,8 901470 | 5651,8 | 395026,2 |
| 32 | 185532,7 | 6,5 | 6,5 | 81300,9 | 16257,6 |
| 17651 | 253 | 17651 3529,7 | 22,1 | 22,1 |

|  |  |  |  |  |  |
|---|---|---|---|---|---|
| 2,4 | 2,4 | 167 | 960122,3 | 6019,6 | 420727,8 |
| 54,9 | 315482,3 | 11 | 11 | 138245 | 27644,7 |
| 46981 | 672 | 46981,4 9394,8 | 58,9 | 58,9 |

The game is solved by the Brown-Robinson method, the value of the game is 35189.49. The evader strategy (1/20, 0, 0, 0, 0, 0, 1/20, 3/10, 3/5), the pursuer strategy (1/4, 1/20, 3/5, 0, 0, 1 /10).

We transform the topic into the search and pursuit between quadrotor UAVs. Modify the topic slightly

Let the distance between the fugitive UAV and the ground be 100 meters, the fugitive UAV selects a speed from the set V^E={8,56,78} as the X-axis speed, and selects from the set α={23,37,82} A value as the Y-axis direction. The maximum speed of the chaser is V^P=100 m/min

Then the fugitive policy set is:

$$(\alpha_1, v_1), (\alpha_1, v_2), (\alpha_1, v_3), (\alpha_2, v_1), (\alpha_2, v_2), (\alpha_2, v_3), (\alpha_3, v_1), (\alpha_3, v_2)$$

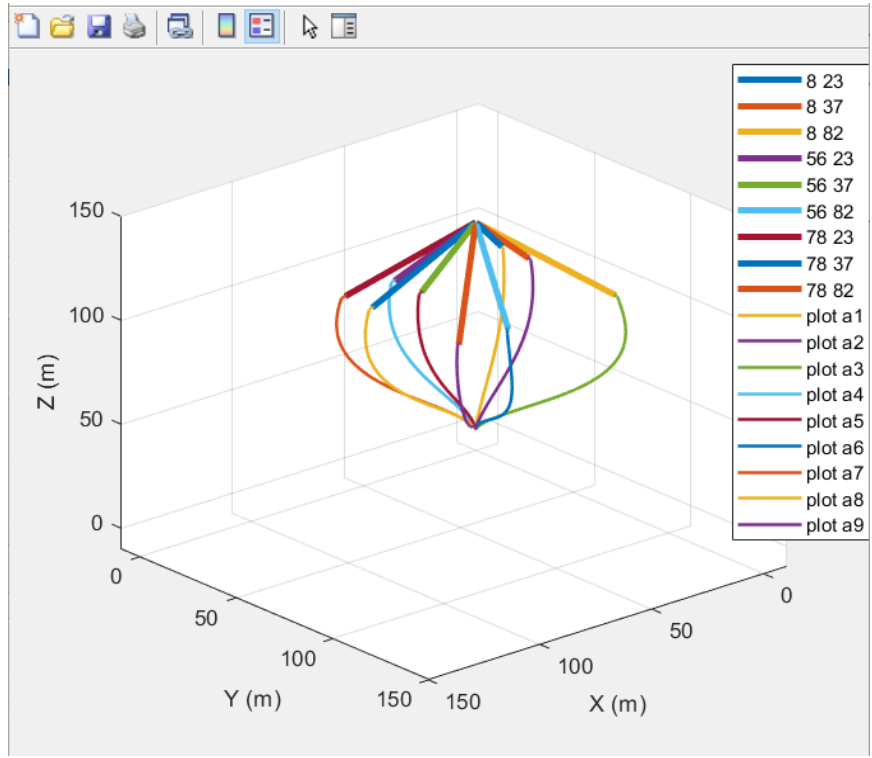

Figure.6. All optional paths and pursuit results of the quadrotor UAV

We transform the topic into the search and pursuit between quadrotor UAVs. Modify the topic slightly

Let the distance between the fugitive UAV and the ground be 100 meters, the fugitive UAV selects a speed from the set V^E={8,56,78} as the X-axis speed, and selects from the set α={23,37,82} A value as the Y-axis direction. The maximum speed of the chaser is V^P=100 m/min

Then the fugitive policy set is:

$$(\alpha_1, v_1), (\alpha_1, v_2), (\alpha_1, v_3), (\alpha_2, v_1), (\alpha_2, v_2), (\alpha_2, v_3), (\alpha_3, v_1), (\alpha_3, v_2)$$

**Example 2**. Let the initial distance between the pursuer and the evader be 50 kilometers. The evader chooses a speed from the set $V^E$= {4,10,16} and a direction from the set α={8,10,16}. The maximum speed of the pursuer is $V^P$=80 km/h. Then the set of strategies for the evader is: $(\alpha_1, v_1), (\alpha_1, v_2), (\alpha_1, v_3), (\alpha_2, v_1), (\alpha_2, v_2), (\alpha_2, v_3), (\alpha_3, v_1), (\alpha_3, v_2)$,, and the set of strategies for the pursuer:

$$(v_1, v_2, v_3), (v_1, v_3, v_2), (v_2, v_1, v_3), (v_2, v_3, v_1), (v_3, v_1, v_2), (v_3, v_2, v_1)$$

The resulting game matrix looks like this:

$$\begin{matrix} 0{,}6 & 0{,}6 & 1{,}325{,}56 & 2{,}161 & 4{,}77 \\ 0{,}9 & 3{,}79 & 0{,}570{,}57 & 3{,}249 & 2{,}039 \\ 2{,}2 & 0{,}9 & 2{,}191{,}38 & 0{,}536 & 0{,}536 \end{matrix}$$

$$\begin{matrix} 0{,}6 & 0{,}6 & 1{,}33\ 5{,}57 & 2{,}165 & 4{,}778 \\ 0{,}9 & 3{,}8 & 0{,}570{,}568 & 3{,}263 & 2{,}048 \\ 2{,}21 & 1 & 2{,}21\ 1{,}39 & 0{,}54 & 0{,}54 \end{matrix}$$

|      |      |           |      |       |       |
|------|------|-----------|------|-------|-------|
| 0,6  | 0,6  | 1,33      | 5,6  | 2,177 | 4,803 |
| 0,92 | 3,86 | 0,58      | 0,576| 3,306 | 2,075 |
| 2,26 | 1,02 | 2,26      | 1,42 | 0,551 | 0,551 |

We transform the topic into the search and pursuit between quadrotor UAVs. Modify the topic slightly

Let the distance between the fugitive UAV and the ground be 100 meters, the fugitive UAV selects a speed from the set $V^E=\{4,10,16\}$ as the X-axis speed, and selects from the set $\alpha=\{8,10,16\}$ A value as the Y-axis direction. The maximum speed of the chaser is $V^P=100$ m/min

Then the fugitive policy set is:

$$(\alpha_1, v_1), (\alpha_1, v_2), (\alpha_1, v_3), (\alpha_2, v_1), (\alpha_2, v_2), (\alpha_2, v_3), (\alpha_3, v_1), (\alpha_3, v_2)$$

The game was solved using the method of Brown-Robinson, and the value of the game is 1.57. The strategy for the evader is (1/20, 0, 0, 0, 0, 0, 1/10, 1/4, 3/5), and the strategy for the pursuer is (9/20, 1/20, 3/20, 1/20, 1/4, 1/20). The solutions of the examples showed that the most probable speed for the evader will be the maximum of the possible speeds. Therefore, the pursuer should start checking the speeds from the maximum possible speed.

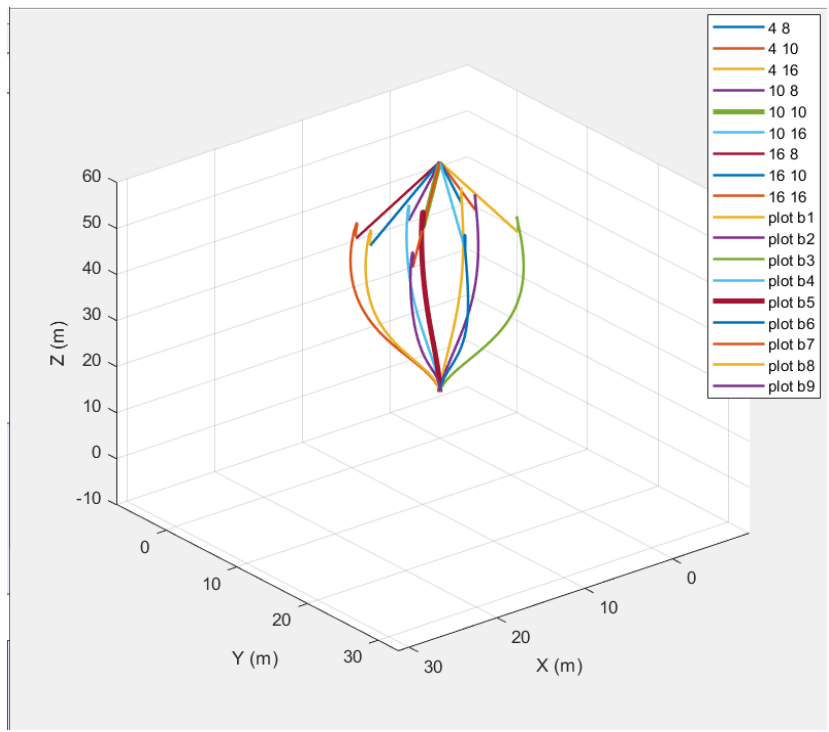

Figure.7.  All optional paths and pursuit results of the quadrotor UAV

## TASK OF ONE PURSUER AND GROUP OF FUGITIVES

Given a set J of n submarines, for which the times of their capture by the pursuer are known, so that |J| denotes the time of capture of J.

For each fugitive, the readiness times of the submarines are also known (possibly times when they will reach a certain place where pursuit cannot continue) $\overline{D}(J)$ and the times required for their execution $\overline{D}(J)$.

For each fugitive, a weight coefficient w(J) is given, which participates in the objective function that needs to be optimized.

The end time of the search is denoted by $t^i$. Thus, $t^i = t_i + T_i$

Examples of criteria functions:
1. Minimize the total penalty for delays.

$$f_1 = \sum_{k=1}^{n-1} w(J_k)(t^k - \overline{D}_k)^+$$

2. Minimize the maximum penalty for delays.

$$f_2 = \max_k w(J_k)(t^k - \overline{D}_k)^+$$

3. Minimize the amount of fines

$$f_3 = \sum_{k=1}^{n-1} w(J_k)(t^k - \overline{D}_k)$$

4. Minimize the amount of tied funds

$x+$ denotes its positive part, defined by the formula $x+=12(x+x)$, Then the delay in catching i of the evaders is equal

$$f_4 = \sum_{k=1}^{n-1} w(J_k)t^k$$

Decision by criterion $f_4$

Let's consider the solution for the function $f_4$ only in the case where $\overline{D}(J) = 0$ for any J∈J. Let's consider the optimal sequence and swap two adjacent elements $J_k$ and $J_{k+1}$ in it. In this case, the capture time of the last one may increase (otherwise the considered solution is not optimal). However, the difference in the criterion function between searching for the first k+1 fleeing members in the modified and optimal order does not exceed.

$$(w(J_{k+1})|J_{k+1}| + w(J_k)(|J_k| + |J_{k+1}|)) - (w(J_k)|J_k| + w(J_{k+1})(|J_k| + |J_{k+1}|)) \geq 0$$

Hence, after reductions, we get.

$$w(J_k)|J_{k+1}| \geq w(J_{k+1})|J_k|$$

Therefore, for the optimal schedule for any k, we obtain the inequality of relations.

$$\frac{|J_k|}{w(J_k)} \leq \frac{|J_{k+1}|}{w(J_{k+!})}$$

Note that if this ratio is equal, the permutation of k+1 and k does not change the value of the criterion. Therefore, any

## CONSIDER THE FOLLOWING SITUATION.

Suppose a intercepting ship, having n boats with depth bombs on board, at time t detects periscopes of n submarines at various distances from it on the surface of the sea, which at the same moment dived underwater and began to move in different directions at fixed speeds. It is required to send the boats to intercept the submarines in an optimal way, that is, so that the sum of the guaranteed times of interception of the submarines would be minimal. To solve the problem, we will create a matrix of efficiency A=(a i j), where each element is the guaranteed time of interception of submarine j by boat i , which consists of the time of reaching the periscope detection point by the boat and its total time of passing along the logarithmic interception spiral. Let $x_{ij}$ be variables that can take only 2 values 0 or 1 as follows.

$$x_{ij} = \begin{cases} 1, assigned\ i\ boat\ for\ j\ submarine \\ 0, assigned\ i\ boat\ for\ j\ submarine \end{cases}$$

It is necessary to find an assignment plan - a matrix X= $\{x_{ij}\}$, i=1...m, j=1...n, which minimizes the search time, while ensuring that each boat is assigned to search for no more than one submarine, and each submarine can be searched by no more than one boat.

Mathematical formulation of the optimal assignment problem.

$$min\ z = min \sum_{i=1}^{m} \sum_{j=1}^{n} \tau_{ij} * x_{ij}$$

$$\sum_{i=1}^{m} x_{ij} \leq 1, j = 1..n$$

$$\sum_{j=1}^{n} x_{ij} \leq 1, i = 1..m$$

$$x_{ij} \geq 0$$

In order for the optimal assignment problem to have an optimal solution, it is necessary and sufficient that the number of boats is equal to the number of submarines, i.e., n=m. Under this condition, the inequality constraints become equality constraints.

$$min\ z = min \sum_{i=1}^{n} \sum_{j=1}^{n} \tau_{ij} * x_{ij}$$

$$\sum_{i=1}^{n} x_{ij} = 1, j = 1..n$$

$$\sum_{j=1}^{n} x_{ij} = 1, i = 1..n$$

$$x_{ij} \geq 0$$

If n≠m, then the assignment problem is unbalanced. Any assignment problem can be balanced by introducing the necessary number of dummy boats or submarines. The dual problem of the optimal assignment problem.

$$\max \omega = \max(\sum_{i=1}^{n} i + \sum_{i=1}^{n} i)$$

$$i + i \geq \tau_{ij}, i = 1..n, j = 1..n$$

The Hungarian method can be used to solve the assignment problem. The essence of the method is as follows:

In the original matrix A of performances, determine the minimum element in each row and subtract it from all other elements in the row.

In the matrix obtained in the first step, determine the minimum element in each column and subtract it from all other elements in the column. If a feasible solution is not obtained after steps 1 and 2, perform:

In the last matrix, draw the minimum number of horizontal and vertical lines through rows and columns to cross out all zero elements.

Find the minimum non-crossed-out element and subtract it from all other non-crossed-out elements and add it to all elements at the intersection of the lines drawn in the previous step.

If the new distribution of zero elements does not allow a feasible solution to be constructed, repeat step 2a. Otherwise, proceed to step 3.

The optimal assignments will correspond to the zero elements obtained in step 2.

Let's consider some numerical examples of solving the problem of distributing boats for catching several submarines.

**Example 3**. Let a interceptor ship detect 4 submarines. The initial distance to each of them is 100 km, 200 km, 50 km, and 163 km, respectively. The pursuer has 4 boats for catching the submarines. The maximum speed of each boat is 74 km/h, 90 km/h, 178 km/h, and 124 km/h, respectively. The first submarine moves along the straight line $\alpha_1$=23, with the speed $v_1$=23 km/h, the second one $\alpha_2$=137, $v_2$=50 km/h, the third one $\alpha_3$=187, $v_3$=67 km/h, and the fourth one $\alpha_4$=50, $v_4$=70 km/h. Then the matrix for the assignment problem looks as follows:

$$\begin{matrix} 1,18 & 0,98 & 0,52 & 0,73 \\ 14,43 & 7,06 & 1,77 & 3,3 \\ 373,78 & 12,12 & 0,77 & 2,13 \\ 14,43 & 3 & 0,96 & 1,53 \end{matrix}$$

We solve the game using the Hungarian method. The value of the objective function is 8.08, the final table looks like this.

$$\begin{matrix} \boxed{0} & 0 & 2,37 & 1,22 \\ 9,63 & 2,46 & \boxed{0} & 0,17 \\ 369,98 & 8,52 & 0 & \boxed{0} \\ 11,22 & \boxed{0} & 0,79 & 0 \end{matrix}$$

We transform the topic into the search and pursuit between quadrotor UAVs. Modify the topic slightly

Suppose an intercepting quadcopter detects 4 intruding quadcopters. Chaser has 4 ships to chase the submarine. The maximum speed of each ship in XYZ axis is 74 km/h, 90 km/h, 178 km/h and 124 km/h respectively.

The first invasion quadrotor UAV, the maximum speed of the X-axis $v\_1=23$m/min, the maximum speed of the Y-axis $\alpha\_1=23$m/min, the height is 100 meters

The second invading quadrotor UAV has a maximum speed of X-axis $v\_2=50$m/min, a maximum speed of Y-axis $\alpha\_2=137$m/min, and a height of 200 meters.

The third invading quadrotor UAV, the maximum speed of the X-axis $v\_3=67$m/min, the maximum speed of the Y-axis $\alpha\_3=7$m/min, and a height of 50 meters

The fourth intrusion quadrotor UAV, the maximum speed of the X-axis $v\_4=70$m/min. Y-axis maximum speed $\alpha\_4=50$m/min, height 163 meters

matching matrix:

$$\begin{matrix} 0 & 0 & 1 & 0 \\ 1 & 0 & 0 & 0 \\ 0 & 1 & 0 & 0 \\ 0 & 0 & 0 & 1 \end{matrix}$$

The value of the objective function is    3.0888

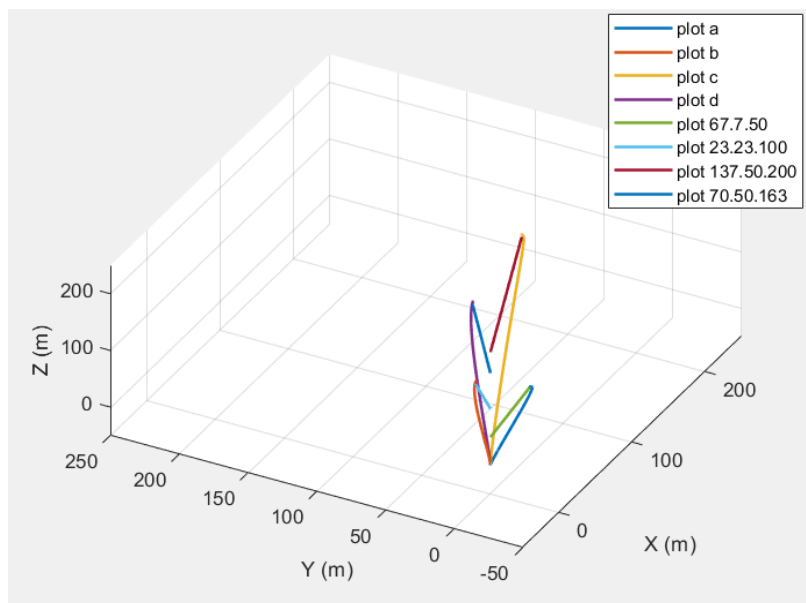

Figure.8. Quadrotor drone matching path

**Example 4**. Let an interceptor ship detect 4 submarines. The initial distance to each of them is 30 km, 11 km, 62 km, and 8

km, respectively. The pursuer has 4 boats for catching the submarines. The maximum speed of each boat is 60 km/h, 65 km/h, 95 km/h, and 105 km/h, respectively. The first submarine moves along the straight line $\alpha_1=7$, with the speed $v_1=7$ km/h, the second one $\alpha_2=11$, $v_2=11$ km/h, the third one $\alpha_3=30$, $v_3=30$ km/h, and the fourth one $\alpha_4=44$, $v_4=44$ km/h. Then the matrix for the assignment problem looks as follows:

$$\begin{matrix} 0,46 & 0,42 & 0,297 & 0,27 \\ 0,16 & 0,15 & 0,11 & 0,097 \\ 0,93 & 0,86 & 0,59 & 0,54 \\ 0,18 & 0,15 & 0,09 & 0,08 \end{matrix}$$

We solve the game using the Hungarian method. The value of the objective function is 1.147, the final table looks like this.

$$\begin{matrix} 0,093 & 0,063 & \boxed{0} & 0 \\ \boxed{0} & 0 & 0,02 & 0,034 \\ 0,29 & 0,23 & 0,023 & \boxed{0} \\ 0,02 & \boxed{0} & 0 & 0,017 \end{matrix}$$

We transform the topic into the search and pursuit between quadrotor UAVs. Modify the topic slightly

Suppose an intercepting quadcopter detects 4 intruding quadcopters. Chaser has 4 ships to chase the submarine. The maximum speed of each ship in XYZ axis is 60 m/min, 65 m/min, 95 m/min and 105 m/min respectively.

The first invasion quadrotor UAV, the maximum speed of the X-axis v_1=7m/min, the maximum speed of the Y-axis α_1=7m/min, the height is 30 meters

The second invading quadrotor UAV has a maximum speed of X-axis v_2=11m/min, a maximum speed of Y-axis α_2=11m/min, and a height of 11 meters.

The third invading quadrotor UAV, the maximum speed of the X-axis v_3=30m/min, the maximum speed of the Y-axis α_3=30m/min, and a height of 62 meters

The fourth intrusion quadrotor UAV, the maximum speed of the X-axis v_4=44m/min. Y-axis maximum speed α_4=44m/min, height 44 meters

matching matrix:

$$\begin{matrix} 0 & 1 & 0 & 0 \\ 0 & 0 & 0 & 1 \\ 1 & 0 & 0 & 0 \\ 0 & 0 & 1 & 0 \end{matrix}$$

The value of the objective function is: 0.8390

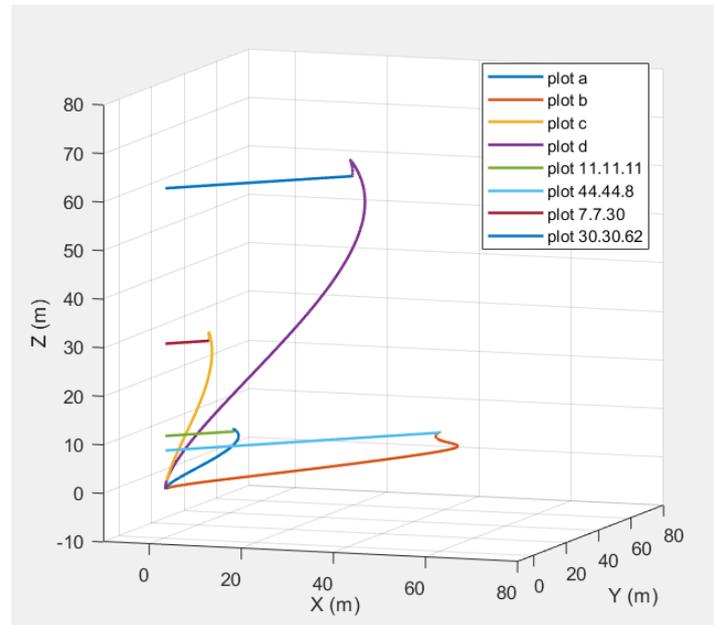

Figure.9. Quadrotor drone matching path

## CONCLUSION

With reasonable parameters of chasing UAVs, considering the success rate and interception efficiency, after calculating all UAV motion parameters using the Hungarian algorithm, all escaped quadrotor UAVs can be chased and successfully intercepted considering the performance parameters such as the speed of each interceptor and the compatibility between the quadrotor UAVs in the interceptor UAV camp and the UAVs in the escaped UAV camp.